\documentclass[usenatbib]{mnras}

\usepackage[T1]{fontenc}
\usepackage{ae,aecompl}

\usepackage{graphicx}	
\usepackage{amsmath}	
\usepackage{amssymb}	
\usepackage{bm}
\usepackage{ulem}
\usepackage{multirow}
\usepackage{balance}
\usepackage{times}

\newcommand{\mP}{{\mathcal{P}}}
\newcommand{\mC}{{\mathcal{C}}}

\newcommand{\vx}{\textbf{x}}
\newcommand{\vr}{\textbf{r}}
\newcommand{\vs}{\textbf{s}}

\newcommand{\vv}{\textbf{v}}

\newcommand{\dd}{\hbox{d}}

\newcommand{\rmHI}{{\rm HI}}

\newcommand{\rmhm}{{\rm HM}}
\newcommand{\rmm}{{\rm m}}
\newcommand{\rmt}{{\rm t}}

\usepackage{color}
\definecolor{Red}{rgb}{0.65,0.08,0.05}
\definecolor{Blue}{rgb}{0.05,0.08,0.65}
\definecolor{Purple}{RGB}{143,51,143}
\definecolor{Green}{rgb}{0.05,0.65,0.05}

\newcommand{\Cora}[1]{#1}

\newcommand{\Ref}[1]{#1}

\title[Counts-in-cells for 21cm intensity mapping]{
Extreme Spheres: Counts-in-cells for 21cm intensity mapping
}

\author[Leicht et al.]{
\parbox[t]{\textwidth}
{\hskip -0.0cm Oliver Leicht$^{1}$\thanks{E-mail: ol248@cam.ac.uk}, Cora Uhlemann$^{1,2}$, Francisco Villaescusa-Navarro$^3$,\\
Sandrine Codis$^4$, Lars Hernquist$^5$, Shy Genel$^{3,6}$}
 \vspace*{6pt}\\
\noindent 
$^{1}$ Centre for Theoretical Cosmology, DAMTP, University of Cambridge, CB3 0WA, United Kingdom\\
$^{2}$ Fitzwilliam College, University of Cambridge, CB3 0DG, United Kingdom\\
$^{3}$ Center for Computational Astrophysics, Flatiron Institute, 162 5th Avenue, 10010, New York, NY, USA\\
$^{4}$ CNRS \& Sorbonne Universit\'e, UMR 7095, Institut d'Astrophysique de Paris, 98 bis Boulevard Arago, 75014 Paris, France\\
$^{5}$ Harvard-Smithsonian Center for Astrophysics, 60 Garden Street, Cambridge, MA 02138, USA\\
$^{6}$ Columbia Astrophysics Laboratory, Columbia University, 550 West 120th Street, New York, NY 10027, USA
}

\date{Accepted XXX. Received YYY; in original form ZZZ}

\pubyear{2018}

\begin{document}
\label{firstpage}
\pagerange{\pageref{firstpage}--\pageref{lastpage}}
\maketitle

\begin{abstract}
Intensity mapping surveys will provide access to a coarse view of the cosmic large-scale structure in unprecedented large volumes at high redshifts. Given the large fractions of the sky that can be efficiently scanned using emission from cosmic neutral hydrogen (HI), intensity mapping is ideally suited to probe a wide range of density environments and hence to constrain cosmology and fundamental physics. To efficiently extract information from 21cm intensities beyond average, one needs non-Gaussian statistics that capture large deviations from mean HI density. Counts-in-cells statistics are ideally suited for this purpose, as the statistics of matter densities in spheres can be predicted accurately on scales where their variance is below unity. 

We use a large state-of-the-art magneto-hydrodynamic simulation from the IllustrisTNG project to determine the relation between neutral hydrogen and matter densities in cells. We demonstrate how our theoretical knowledge about the matter PDF for a given cosmology can be used to extract a parametrisation-independent HI bias function from a measured neutral hydrogen PDF. When combining the predicted matter PDFs with a simple bias fit to the simulation, we obtain a prediction for neutral hydrogen PDFs at a few percent accuracy at scale $R=5$ Mpc$/h$ from redshift $z=5$ to $z=1$. Furthermore, we find a density-dependent HI clustering signal that is consistent with theoretical expectations and could allow for joint constraints of HI bias and the amplitude of matter fluctuations or the growth of structure.

\end{abstract}



\section{Introduction}
Upcoming large-scale, post-reionisation intensity mapping surveys like Tianlai \citep{Chen_2012}, BINGO \citep{Battye13BINGO}, CHIME \citep{CHIME14pathfinder}, FAST \citep{fast},  HIRAX \citep{Newburgh16HIRAX}, MeerKAT \citep{MeerKAT}, SKA \citep{Santos15SKA} and SPHEREx \citep{Dore16SPHEREx} will sample the spatial distribution of cosmic matter through tracers of it, such as neutral hydrogen, at redshifts $0<z<6$. The advantages of those surveys with respect to traditional optical methods to map galaxies is that they can sample very large cosmological volumes in a very efficient manner.
Following early ideas of intensity mapping \citep{Madau97,Bharadwaj01,Battye04,Barkana05}, the first detection of the 21cm cosmological signal was achieved by cross-correlating 21cm intensity maps from the Green Back Telescope with the DEEP2 optical galaxy survey \citep{Pen09,Chang10,Masui13}. While we have not yet detected the 21cm cosmological signal in auto-correlation in the post-reionisation era\footnote{see \cite{Bowman} for a detection claim at high-redshift.} \citep{Switzer13}, upcoming surveys will have sensitivity enough to allow us to study cosmology at an unprecedented precision with both auto- and cross-correlations \citep{Bull15,Pourtsidou17,Kovetz17,Paco_15, Paco_14, Carucci_15, Andrej_17, Carucci_17,Paco_Lyb}.
Furthermore, we can probe dark energy through baryonic acoustic oscillations \citep{Chang08, Paco_17} or approach weak lensing of intensity mapping \citep{Harrison_16, Bonaldi_16, Foreman_18, Schaan18} by using the background as a source image.

It is well known that the nonlinear evolution of matter in the Universe introduces a leakage of information from the two-point correlation function (or the power spectrum) into higher-order terms \citep{Scoccimarro_99}. Thus, in order to extract the maximum information from large-scale structure surveys at low redshifts, we need to consider quantities beyond the two-point correlation or to attempt to reconstruct the linear fields \citep{Schmittfull_15}. In this work, we focus on the former approach and consider one-point statistics as complementary source of cosmological information compared to traditional two-point statistics. 

For the epoch of reionisation, one-point statistics and higher-order moments have been proposed as sources of information about the physics of reionisation and the nature of ionising sources \citep{Harker09,Ichikawa10,Baek10,Shimabukuro15,Kittiwisit18}. At later times, counts-in-cells statistics can capture essential non-Gaussian information from the 21cm intensity (and hence HI density) field that is lost in common two-point statistics and add information about the density-dependence of clustering. Furthermore, the underlying matter statistics in real space can be analytically predicted \citep{Bernardeau2014,Uhlemann16log} from first principles and at percent accuracy for scales at which the variance of the smoothed matter density is below unity. Those scales are typically above $10$ Mpc$/h$ at redshift $z=0$, such that the typical low-angular resolution inherent to intensity mapping is not a major limiting factor. 

The formalism, based on large-deviation statistics, allows us to access the rare event tails probing large density fluctuations that  contain valuable information about fundamental physics (such as neutrino masses, primordial non-Gaussianity and modified gravity) that are inaccessible to common perturbative methods. To tap the potential of this probe for cosmology, we build upon a previous study of dark matter halos \citep{Uhlemann18bias} and quantify the nonlinear bias function that relates matter and neutral hydrogen counts-in-cells on scales where it is nonlinear and distinct from the bias measured from two-point clustering \citep{Castorina17,Villaescusa-Navarro18}. While we focus on cosmology here, counts-in-cells statistics are also used to constrain important astrophysical ingredients such as luminosity functions \citep{Breysse17}. Those could potentially be improved by predictions from large-deviation statistics, which are more accurate than phenomenological lognormal models that are currently used.

We use the magneto-hydrodynamic simulation IllustrisTNG simulation to compare counts-in-cells of neutral hydrogen, halos (as galaxy proxies) and matter. Extending on recent results for the clustering statistics of neutral hydrogen \citep{Villaescusa-Navarro18}, we quantify the counts-in-cells bias between the different tracers in IllustrisTNG and determine the effect of redshift-space distortions. Based on this, we assess the promise of mock catalogs to efficiently access exquisitely sampled counts-in-cells while mitigating inaccuracies from neglecting the 1-halo term and Fingers-of-God that appeared as deal breakers for the power spectrum.

This paper is organised as follows:  
Section~\ref{sec:IllustrisTNG} describes the IllustrisTNG simulation and how we extracted counts-in-cells statistics. In Section~\ref{sec:DM} we briefly recap the theoretical formalism that allows us to obtain the probability distribution function (PDF) and the density-dependent correlation of matter densities in spheres. Section~\ref{sec:tracer} discusses how one can relate these results to the tracer PDF and density-dependent clustering using bias models. We present the results for the bias relation between matter and neutral hydrogen along with the combined predictions for the neutral hydrogen PDF and density-dependence of clustering in Section~\ref{sec:results}. Section~\ref{sec:Conclusion} presents our conclusions and provides an outlook of the potential applications of our findings.

\begin{figure}
\includegraphics[width=\columnwidth]{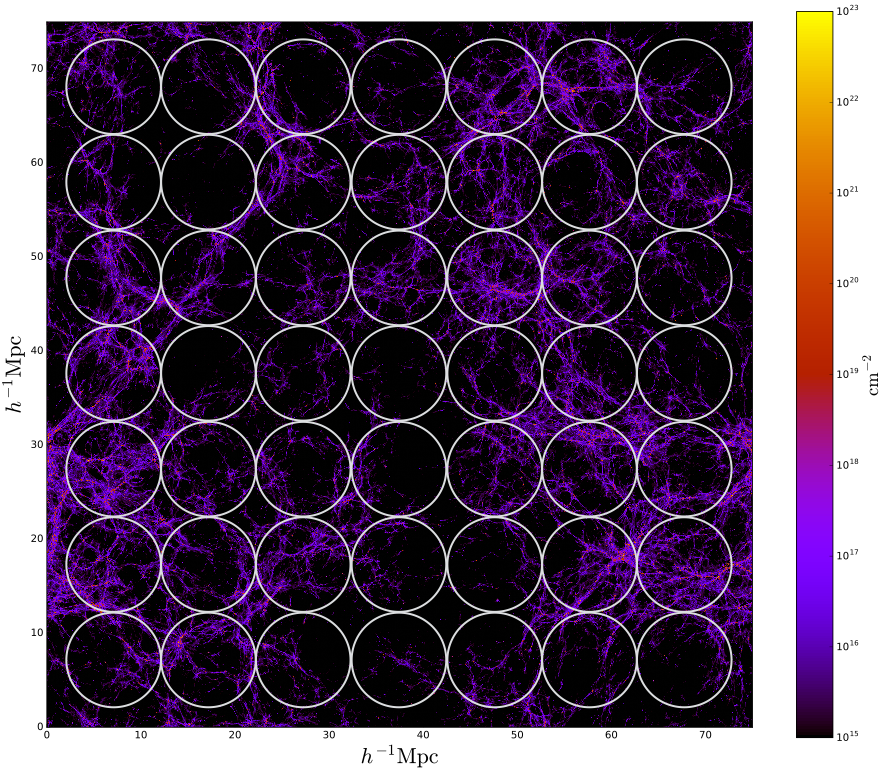}
\caption{Spatial distribution of neutral hydrogen in a $5$ Mpc$/h$ slice of the TNG100 simulation at redshift $z=3$. We also show a small subset of the spheres with radius $R=5$ Mpc$/h$ used for counts-in-cells statistics.}
   \label{fig:IllustrisTNG}
\end{figure}

\section{Numerical simulation}
\label{sec:IllustrisTNG}
The simulation used in this work is part of the IllustrisTNG project \citep[see][where stellar mass and assembly, clustering, colours, magnetic fields and chemical enrichment are discussed]{PillepichA_17a,SpringelV_17a,NelsonD_17a,MarinacciF_17a,NaimanJ_17a}. 
We employ here the TNG100 cosmological box
(containing the same volume as the original Illustris simulation; \citealp{VogelsbergerM_14a,VogelsbergerM_14b,GenelS_14a}) that has been evolved down to $z=0$,
with a comoving box side length of $75$ Mpc$/h$. 

The simulation was run with the {\small AREPO}
code \citep{Arepo}, which evolves the initial conditions accounting for gravity (using a TreePM method), magneto-hydrodynamics \citep[through a Godunov approach on a moving Voronoi mesh]{Pakmor11}
and a range of astrophysical processes described by subgrid
models. These processes include primordial and metal-line cooling,
assuming a time-dependent uniform UV background radiation, star and
supermassive black hole formation, stellar population evolution that
enriches surrounding gas with heavy elements or metals, galactic winds, and
several modes of black hole feedback. 
The numerical methods and subgrid physics models are an update of the Illustris galaxy formation model \citep{Vogelsberger13,Torrey14} and specified in detail
in \cite{WeinbergerR_16a} and \cite{Pillepich_2017b}. Importantly, where uncertainty
and freedom exist for the implementation of these subgrid models, they are
parametrized and tuned to obtain a reasonable match to a small set of
observational properties \citep{PillepichA_17a}. These include the galaxy stellar mass function, the stellar-to-halo mass relation, and the stellar size-mass relation, all at $z=0$.
Additional aspects including feedback effects of black holes, the evolution of galaxy and halo sizes and metallicities are discussed in \cite{Weinberger_2017,GenelS_17a,VogelsbergerM_17a} and \cite{TorreyP_18a}.

\paragraph*{HI modeling.}
We model the spatial distribution of neutral hydrogen accounting for ionisation equilibrium with the UV background, HI self-shielding and the presence of molecular hydrogen following the method depicted in \cite{Villaescusa-Navarro18}, to which we refer the reader for further details. For a study on the impact of the H$_2$ model on HI properties see \cite{Diemer18}.

\paragraph*{Halo identification.}
In this paper we work with halos identified by the Friends-of-Friends (FoF) algorithm with a linking length of
$b=0.2$ \citep{DavisM_85a}. The halo center is identified as the position of
the most bound particle in the halo.  The minimum halo mass we consider is around $2\times 10^8 M_\odot/h$.

\paragraph*{Counts-in-cells.}
We extracted counts-in-cells (i.e. mean densities in spheres) of the matter, neutral hydrogen and mass-weighted halo field in overlapping spheres of comoving radius $R=5$ Mpc$/h$ on a regular grid of size $128^3$ yielding approximately 2 million density samples. Figure~\ref{fig:IllustrisTNG} shows a snapshot of the neutral hydrogen distribution in the TNG100 simulation at redshift $z=3$ along with a subsample of the spheres used for our counts-in-cells analysis. When we write $\rho_\rmm$, $\rho_\rmHI$ and $\rho_{\rm HM}$, we refer to the density in spheres (hence smoothed at radius $R$) of matter, neutral hydrogen and mass-weighted halos, respectively. All densities are measured in units of mean density and are hence dimensionless and related to the density contrast $\delta$ as $\rho=1+\delta$. Note that when measuring this for discrete tracers such as galaxies that are a coarse and biased sampling of the underlying field, one needs to include shot-noise contributions, for example through Poisson sampling, see e.g. \cite{Szapudi2004,Friedrich17}. The objects considered in this work are expected to have very low shot-noise amplitudes\footnote{The very high-resolution of the IllustrisTNG simulation guaranties a very low shot-noise amplitude for the matter field. \cite{Castorina17, Villaescusa-Navarro18} have shown that the amplitude of the shot-noise is negligible for neutral hydrogen in the post-reionisation era. Mass-weighted halos are also expected to have a very low shot-noise amplitude \citep{Seljak_09}.}, so we neglect, for simplicity, their contribution to counts-in-cells.

The PDFs, encoding the probability of finding a certain density in a randomly drawn sphere of fixed radius, were estimated using kernel density estimation.
In cases where a discretisation is needed, we use histograms with a logarithmic binning in densities such that each bin contains approximately $1/75$ of the probability mass. Error bars at those sampling points are determined by means of a jackknife estimator with 30 random subsamples. The random selection ensures more independent spheres within each subsample in comparison to splitting the volume into 30 regular subboxes. It mitigates effects from long range modes too, even though there will still be super-sample variance effects due to the small box volume. The random selection might underestimate the error bars, as the different subsamples are correlated. Due to the small cosmological box with a side length of only 75 Mpc$/h$, we chose spheres of radius $R=5$ Mpc$/h$ that are about a factor 2 smaller than the scales one would want to probe at very low redshifts to ensure a variance below unity. Since the shape of the counts-in-cells PDF is driven by the nonlinear matter variance $\sigma^2(R,z)$ at radius $R$ and redshift $z$, which is related to the nonlinear power spectrum according to Eq. \eqref{eq:nonlinvariance}, our formalism can be used to relate different radii and redshifts; for example, the PDF for $R=10$ Mpc$/h$ at $z=0$ closely corresponds to that for $R=5$ Mpc$/h$ at $z=1$, as the amplitude of fluctuations is almost the same in those two cases. 

\paragraph*{Redshift-space mapping.}
To assess the impact of redshift-space distortions, a mapping from real-space to redshift-space was done by converting the comoving positions (of matter, halos and neutral hydrogen) $\vr$ to the redshift-space ones, $\vs$, by shifting them along the fictitious line-of-sight (chosen in x-direction here) according to their peculiar velocity along that direction
\begin{align}
\vs = \vr+ \frac{1+z}{H(z)} \vv\cdot{\hat \vx} \,.
\end{align}

\section{Statistics of matter densities in spheres}
\label{sec:DM}
Before going to tracers in the next section, let us start with the statistics of the matter density field that is the first building block in our modelling of the statistics of neutral hydrogen densities.
\subsection{One-point PDF of matter density}
As shown in \cite{Bernardeau2014}, the PDF for matter densities $\rho_\rmm$ within a sphere of radius $R$ at redshift $z$, $\mP_R(\rho_\rmm)$, valid at mildly nonlinear scales where $\sigma(R,z)\lesssim 1$, can be obtained from large-deviation statistics, that we quickly outline here. The basic idea is to use the fact that the initial PDFs of densities in spheres are Gaussian
\begin{align}
 \label{eq:saddlePDFlog-ini}
\mP_{R_{\rm ini}}^{\rm{ini}}(\delta_L)&= \frac{1}{\sqrt{2 \pi \sigma^2_L(R_{\rm ini})}}  \exp\left[-\frac{\delta_L^2}{2\sigma^2_L(R_{\rm ini})}\right] \,,
 \end{align}
where $\delta_L$ is the initial linear density contrast, and the linear variance $\sigma_L^2$ is determined from the initial power spectrum $P_L$ using the Fourier transform of the spherical top-hat filter $W$ of radius $r$
\begin{align}
\label{eq:linvariance}
\sigma_L^2(r)=\int \frac{\dd^3k}{(2\pi)^3}\, P_L(k) W^2(kr)\,.
\end{align}
Due to the exponential decay of the PDF with increasing linear density contrast $\delta_L$ or decreasing variance, one can seek to identify the most likely dynamics (amongst all possible mappings between the initial and final densities) that can bring about a final density with a possibly large deviation from the mean (a rare event). When considering a highly symmetric observable such as the PDF of density in spheres, one can argue that the most likely gravitational dynamics respects the symmetry \citep{Valageas02} and is hence given by the spherical collapse solution. Then, one can infer the exponential decay of the final density PDF by plugging in the spherical collapse mapping between initial and final densities $\delta_L=\delta_{L,\rm SC}(\rho_\rmm)$ and using mass conservation to relate the initial (Lagrangian) and final (Eulerian) radii of spheres containing the densities $R_{\rm ini}=R\rho_\rmm^{1/3}$. An accurate approximation to spherical collapse dynamics, introduced by \cite{Bernardeau92}, reads
\begin{equation}
\rho_{\rm SC}(\delta_L)\simeq \left (1-\delta_L/\nu\right )^{-\nu} \, \Leftrightarrow\,
\delta_{L,\rm SC}(\rho)\simeq \nu(1-\rho^{-1/\nu})\,,
\label{eq:spherical-collapse}
\end{equation}
where the $\nu$ parametrises the dynamics of spherical collapse. Here we choose $\nu=21/13$ to exactly match the high-redshift skewness obtained from perturbation theory
\citep{Bernardeau2014}.

While the proper derivation of the PDF including its prefactor requires more steps\footnote{While in general, the PDF is obtained from an inverse Laplace transform that requires an integration in the complex plane, \cite{Uhlemann16log} has shown that one can perform an analytical saddle-point approximation in the $\log$-density to obtain a closed form expression that is valid for a wide range of densities.}, one can obtain a simple analytical form by extrapolating the zero-variance limit to small values of the variance \citep{Uhlemann16log}
\begin{subequations}
\label{eq:PDFfromPsi}
\begin{equation}
\hskip -0.1cm \mP_R(\rho_\rmm) \!=\! \sqrt{\frac{\Psi''_{R}(\rho_\rmm)+\Psi'_{R}(\rho_\rmm)/\rho_\rmm}{2\pi \sigma^{2}_\mu}} \exp\left(-\frac{\Psi_R(\rho_\rmm)}{\sigma^{2}_\mu}\right ),
\label{eq:PDFfromPsi2}
\end{equation}
where the prime denotes a derivative with respect to $\rho_\rmm$ and the exponential decay of the PDF is given by the function
\begin{equation}
\Psi_R(\rho_\rmm)= \frac{\delta^{2}_{L,\rm SC}(\rho_\rmm) \sigma_L^{2}(R)}{2\sigma_L^2(R\rho_\rmm^{1/3})}\,.
\label{eq:Psiquad}
\end{equation}
Here $\delta_{L,\rm SC}(\rho_\rmm)$ is the linear density contrast, averaged within the initial Lagrangian radius $R_{\rm ini}=R\rho_\rmm^{1/3}$, which can be be mapped to the nonlinearly evolved density $\rho_\rmm$ within radius $R$ using the spherical collapse model~\eqref{eq:spherical-collapse}, $\sigma_L^2$ is the linear variance from equation~\eqref{eq:linvariance}, and 
$\sigma_\mu^2\equiv\sigma_\mu^2(R,z)$ is the nonlinear variance of the log-density (because the formula has been derived from an analytic approximation based on the log-density $\mu_\rmm=\log\rho_\rmm$).
To ensure a unit mean density and the correct normalization of the PDF, one has to evaluate the PDF obtained from equation~\eqref{eq:PDFfromPsi2} according to
\begin{align}
\label{eq:PDFfromPsi2norm}
\hat\mP_R(\rho_\rmm)= \mP_R\left(\rho_\rmm \, \frac{\langle\rho_\rmm\rangle}{\langle 1\rangle}\right) \cdot \frac{\langle\rho_\rmm\rangle}{\langle 1\rangle^2} \,,
\end{align}
\end{subequations}
with the shorthand notation $\langle f(\rho_\rmm)\rangle=\int_0^\infty \dd\rho_\rmm\, f(\rho_\rmm)\mP_{R}(\rho_\rmm) $. This step is necessary as equation~\eqref{eq:PDFfromPsi2} ensures the correct tree-level cumulants of order 3 and above, the right nonlinear variance of $\mu_\rmm$ and zero mean for $\mu_\rmm$. Since we, instead, want the density $\rho_\rmm$ to have unit mean, it is necessary to correct for the non-zero value of the mean of $\mu_\rmm$ using equation~\eqref{eq:PDFfromPsi2norm}.

\begin{figure}
\includegraphics[width=\columnwidth]{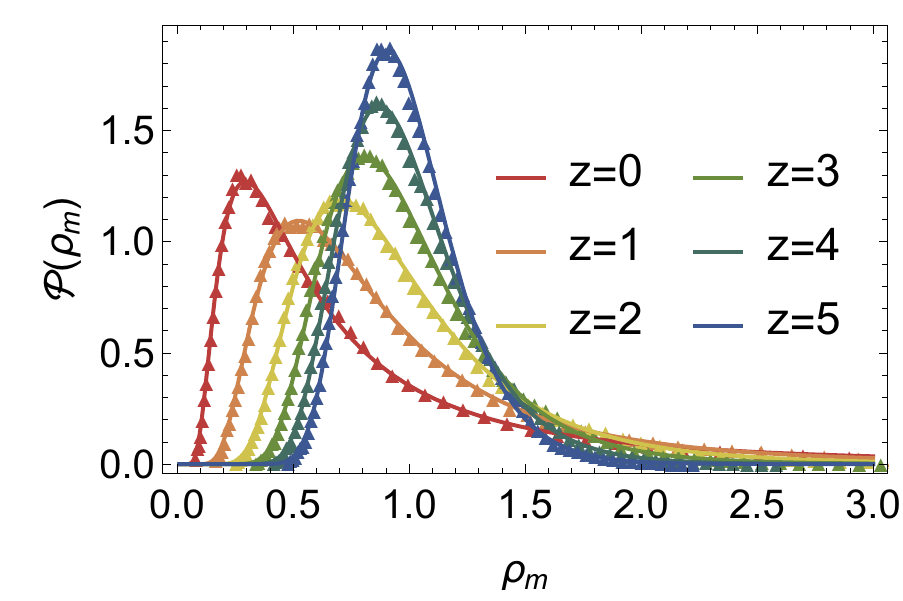}\\
\includegraphics[width=\columnwidth]{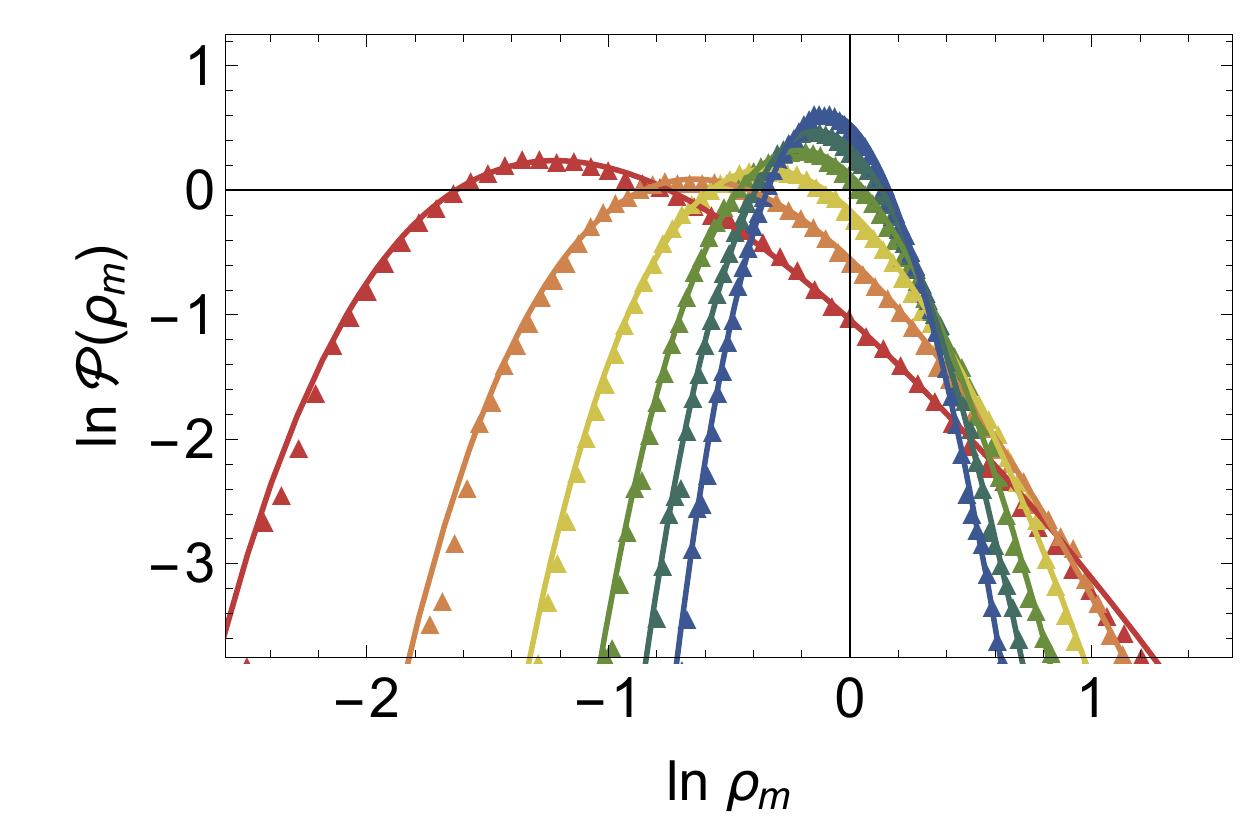}\\
\includegraphics[width=\columnwidth]{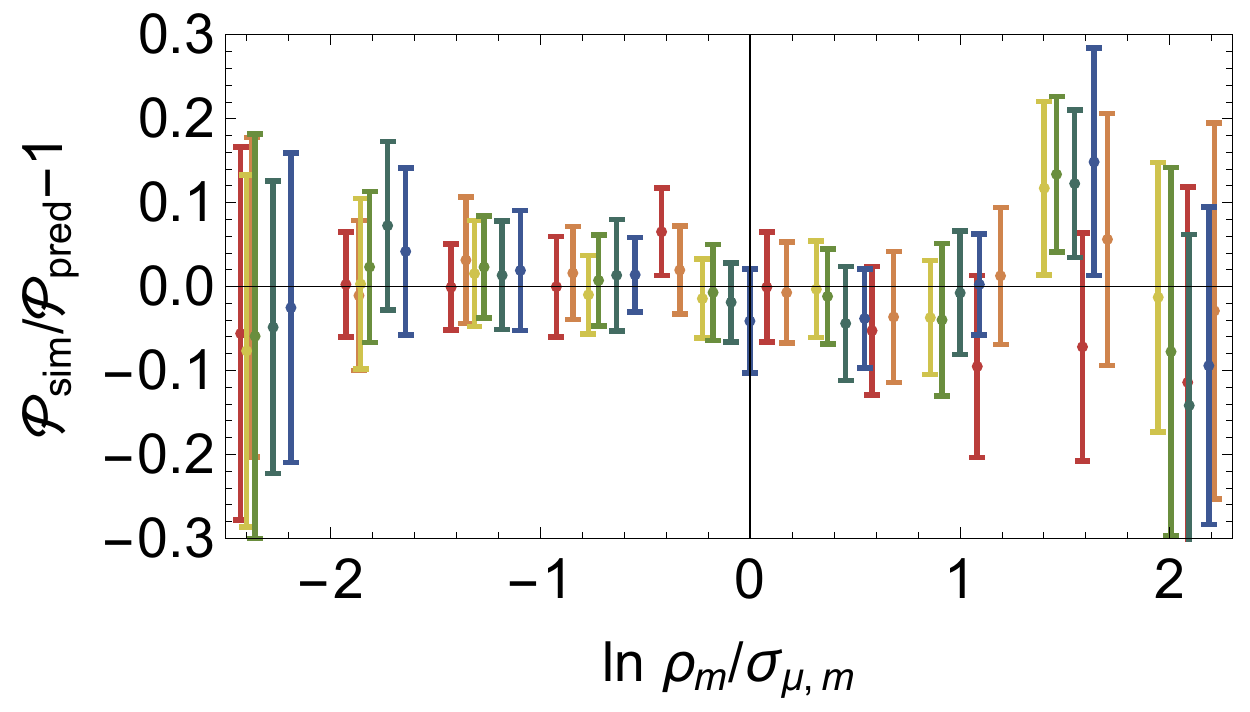}\\
\caption{{\bf Upper and middle panels:} Measured PDF of matter densities in spheres of radius $R=5$ Mpc$/h$ at redshifts $z=0,1,2,3,4,5$ (red to blue data points) compared to the prediction from large-deviation statistics with the measured nonlinear variance as input (dashed lines) in linear scale (upper panel) and log-scale (middle panel). {\bf Lower panel:} Residuals between the theoretical predictions with measured variance and the measured PDFs.}
   \label{fig:DMPDF}
\end{figure}

 \cite{Uhlemann16log} showed that the above model for the PDF of the real-space matter density field $\hat \mP_R(\rho_\rmm|\sigma_\mu)$ with the variance of the log-density $\sigma^2_\mu(R)$ as a driving parameter was accurate at the percent level for standard deviations $\sigma_\mu\lesssim 0.5$, greatly improving over the accuracy that can be obtained from lognormal models \citep{ColesJones91}. To give an impression of the exquisite accuracy, we show a comparison between the measurement from the IllustrisTNG simulation and the theoretical prediction for the matter PDF with the measured nonlinear variance $\sigma_\mu^2$ in Figure~\ref{fig:DMPDF}. As expected, the PDF is close to Gaussian at high redshift and becomes more and more skewed at lower redshifts, as voids are occupying most of the volume and peaks are exceedingly accreting matter. 

When aiming at a fully theoretical model for the matter PDF, one can use the nonlinear variance as predicted by a halofit power spectrum $P$ \citep{Smith2003,halofit14}, in analogy to the linear variance from equation~\eqref{eq:linvariance}, 
\begin{align}
\label{eq:nonlinvariance}
\sigma^2(r)=\int \frac{\dd^3k}{(2\pi)^3}\, P(k) W^2(kr)\,.
\end{align}
The nonlinear variance of the log-density $\sigma_\mu^2$ that enters the PDF is then chosen such that the variance of the PDF in equation~\eqref{eq:PDFfromPsi} matches the nonlinear density variance $\sigma^2$ from halofit.
The halofit nonlinear variance agrees with the measured value to typically better than 1\% which propagates to an additional 1-2\% error on the PDF. At $z=0$, our method of estimating the logarithmic variance from halofit no longer works and causes significant discrepancies. This is not unexpected, given that we are probing smaller scales than it would be desirable at this redshift, and on smaller scales baryonic effects become important, which are not captured in the fit to N-body simulations. Note that, in general one can treat $\sigma_\mu$ as a free parameter of the theory, and we only rely on an approximate predicted value of $\sigma_\mu$ to infer the functional form of the bias relation between matter and neutral hydrogen.

\begin{figure}
\includegraphics[width=\columnwidth]{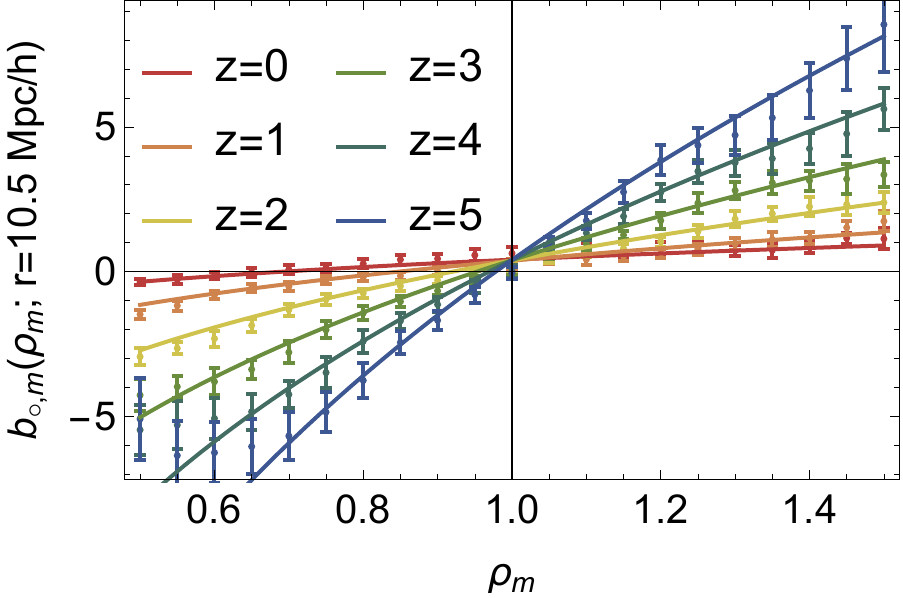}
\caption{Density-dependent clustering encoded in the sphere bias function from equation~\eqref{eq:def2ptbias} for matter at redshifts $z=0$ to $5$ (red to blue) as measured in the IllustrisTNG simulation (data points) and predicted by large-deviation statistics using equation~\eqref{eq:spherebias} (lines).}
   \label{fig:spherebiasDM}
\end{figure}

\subsection{Density-dependent clustering of matter}
Apart from the one-point statistics of density in spheres, one can also extract a density-dependent clustering signal that quantifies
the difference in clustering of regions with high/low densities compared to regions with average densities. This is encoded in the density-dependent correlation function, which is the ratio of \Cora{the joint} two-point PDF of matter densities at separation $r$ and the marginal one-point PDFs
\begin{align}
\label{eq:densitydepcorrelation}
\xi_{\circ,\rmm}(\rho_\rmm,\rho_\rmm',r)  &=\frac{\mP_{R}(\rho_\rmm,\rho_\rmm';r)}{\mP_{R}(\rho_\rmm)\mP_{R}(\rho_\rmm')}-1.
\end{align}
At large separation $r \gg 2R$, this correlation function factorises into an average separation-dependent correlation $\xi_{\circ,\rmm}(r)$ separation-dependent, and a density-dependent modulation called sphere bias $b_{\circ,\rmm}(\rho_\rmm)$ \citep{Codis2016twopoint} 
\begin{align}
\xi_{\circ,\rmm}(\rho_\rmm,\rho_\rmm',r)&\simeq \xi_{\circ,\rmm}(r)b_{\circ,\rmm}(\rho_\rmm)b_{\circ,\rmm}(\rho_\rmm') .
\end{align}
The mean sphere two-point correlation, $\xi_{\circ,\rmm}(r)$, is the standard matter correlation function of densities smoothed at the sphere radius $R$.
The sphere bias measures the excess correlation induced by a matter density $\rho_\rmm$ at separation $r$ and therefore can be defined as a ratio of the conditional mean of the sphere density $\rho_\rmm'$ given a density $\rho_\rmm$ at separation $r$ and the average correlation
\begin{align}
\label{eq:def2ptbias}
b_{\circ,\rmm}(\rho_\rmm)&=\frac{\langle\rho_\rmm'|\rho_\rmm;r\rangle-1}{\xi_{\circ,\rmm}(r)}\ ,\\ 
\notag \xi_{\circ,\rmm}(r)&=\langle\rho_\rmm(\vx)\rho_\rmm(\vx+\vr)\rangle-1\,.
\end{align}
At large separation, the sphere bias becomes independent of separation $r$ and in general can be predicted more accurately than the approximation for the full two-point clustering~\eqref{eq:densitydepcorrelation}. Again, it can be computed using large-deviation statistics with spherical collapse and a rescaling to the nonlinear variance \citep{Bernardeau96,AbbasSheth07,Codis2016twopoint,Uhlemann17Kaiser}
\begin{subequations}
\label{eq:spherebias}
\begin{equation}
\label{eq:spherebiaspre}
b_{\circ,\rmm}(\rho_\rmm)=\frac{\delta_{L,\rm SC}(\rho_\rmm)\sigma_L^2(R)}{\sigma_L^2(R\rho_\rmm^{1/3})\sigma_\mu^2}\,,
\end{equation}
with a normalisation according to
\begin{equation}
\label{eq:spherebiasnorm}
\hat b_{\circ,\rmm}(\rho_\rmm)=\frac{b_{\circ,\rmm}(\rho_\rmm)-\langle b_{\circ,\rmm}(\rho_\rmm)\rangle}{\langle (\rho_\rmm-1)b_{\circ,\rmm}(\rho_\rmm)\rangle} \,.
\end{equation}
\end{subequations}
The validity of equations~\eqref{eq:spherebias} has been established before for dark matter in large cosmological simulations \citep{Uhlemann17Kaiser,Uhlemann18bias} and found to be surprisingly accurate even for separations $r\gtrsim 2R$, that are only slightly larger than twice the sphere radius.

Due to the small box size of TNG100, it is difficult to get reliable measurements of the correlations in the large separation regime,  such that we have to rely on a separation that just ensures non-overlapping spheres. Figure~\ref{fig:spherebiasDM} displays such a measurement of the density-dependent clustering signal for matter densities in real-space, which should be interpreted with caution (given that even the minimal  separation of non-touching spheres is more than one tenth of the box size), but is completely consistent with the theoretical expectation and previous measurements in large N-body simulations. As expected, spheres of particularly large or small densities are more strongly clustered than average densities. When plotted as a function of the density, the clustering appears stronger at high redshifts, because the matter variance is smaller and hence the relative density contrast is larger than at low redshifts.

\section{Statistics of tracer densities in spheres}
\label{sec:tracer}
Let us now turn to biased tracers of matter and describe how the previous results for matter densities in spheres can be mapped to tracer densities in spheres.

\subsection{One-point PDF of tracer density}
\label{sec:HIbias}
 In general, one can express the respective one-point PDFs of matter and the tracer as marginals of their joint one-point PDF $\mP(\rho_\rmm,\rho_{\rm t})$
\begin{equation}
\label{eq:marginals}
\mP_{\rm t}(\rho_{\rm t}) = \!\int\!\dd\rho_\rmm\, \mP(\rho_\rmm,\rho_{\rm t})\,,\ 
\mP_{\rm m}(\rho_{\rm m}) = \!\int\!\dd\rho_{\rm t}\, \mP(\rho_\rmm,\rho_{\rm t})\,.\!
\end{equation}
For simplicity, our bias model is formulated between matter and tracer densities (such as neutral hydrogen, halos or galaxies) for spheres of identical radii.
While this is a simplistic approximation, as in general the relationship is non-local in both space and time \citep[for a review see][]{Desjacques18review}, we will show that it is sufficient for our purpose.

The tracer density PDF, $\mP_{\rm t}$, can be hence written as a convolution of the matter PDF, $\mP_\rmm$, and the conditional PDF of finding a certain tracer density given a matter density 
\begin{equation}
\mP_{\rm t}\left (\rho_{\rm t}\right ) = \int\!\dd\rho_\rmm\, \mP_{\rm bias}(\rho_{\rm t}|\rho_\rmm)\mP_\rmm (\rho_\rmm) ,
\label{eq:CONVOLVE}
\end{equation}
where $\mP_{\rm bias}(\rho_{\rm t}\vert\rho_\rmm)$ is the conditional PDF (i.e. the probability of having a tracer density $\rho_{\rm t}$ given a matter density $\rho_{\rm m}$). This conditional depends on the details of tracer formation and its associated parameters such as halo mass, smoothing scales, redshift and environment, but also includes scatter around any deterministic relation (stochasticity) which results from an incomplete understanding of the formation process. While in principle one could think that the full joint PDF is needed, one can separate this information into the marginals, the one-point PDFs which are of interest here, and correlations between matter and tracer densities that are independent of the marginal PDFs \Ref{(Uhlemann et. al. in preparation)}.

In the following, we will focus on the marginals and determine an accurate mean bias relation that allows for a one-to-one relation between the matter and tracer PDF. This is in the spirit of large-deviation statistics, which argues that the mean local gravitational evolution given by spherical collapse is adequate to predict the PDF of matter densities in spheres\footnote{The large-deviation principle states that the statistics is dominated by the path that minimises the ``action'' -- or in our case the exponential decay of the PDF -- in order to maximise the probability. This most likely path or dynamics can be decomposed into a gravitational part, given by the spherical collapse, and an astrophysical part, given by the mean bias relation.}. Equipped with a bias model for the mean relation $\rho_\rmm(\rho_{\rmt})$, the tracer PDF $\mP_{\rm t}$ is now obtained from the matter PDF $\mP_\rmm$ by conservation of probability
\begin{equation}
\mP_{\rm t}\left (\rho_{\rm t}\right ) = \mP_\rmm(\rho_\rmm (\rho_{\rm t})) \left\lvert \dd\rho_\rmm/\dd\rho_{\rm t}\right\rvert\,,
\label{eq:tracerPDF}
\end{equation}
where it is required that $\rho_\rmm(\rho_{\rm t})$ is a strictly monotonic function. 
  
\subsection{Density-dependent clustering of tracers}
Using a mean bias model, one can also relate the density-dependent clustering of matter in equation~\eqref{eq:spherebias} to a tracer 
\begin{equation}
\label{eq:spherebiastracer}
b_{\circ,\rmt}(\rho_\rmt) = b_{\circ,\rmm}\left(\rho_\rmm(\rho_\rmt)\right) \sqrt{\xi_{\circ,\rmm}/\xi_{\circ,\rmt}} \,,
\end{equation}
\Cora{where the ratio of correlation functions can be computed as}
\begin{equation}
\label{eq:ratiocorr}
\sqrt{\xi_{\circ,\rmm}/\xi_{\circ,\rmt}} =\left\langle \rho_{\rmt}(\rho_{\rmm}) b_{\circ,\rmm}(\rho_{\rmm})\right\rangle \,.
\end{equation}

\subsection{Parametrisation-independent bias functions}
Previously, we have seen that the question of how to obtain an accurate model for the statistics of tracer densities in spheres boils down to successfully describing the effective mean bias relation between matter and the corresponding tracer densities in spheres. 

The advantage of obtaining bias functions in a parametrisation-independent way is that they can be used as guiding principle for finding suitable parametrisations with a small number of parameters that capture their functional form. This is particularly important if one is interested in the tails of the distribution where common polynomial bias models do not lead to satisfactory results.
Since we want to  map the matter PDF to the tracer PDF,  let us rely on an `inverse' bias model $\rho_\rmm(\rho_{\rm t})$ writing the matter density as a function of the tracer density rather than the other way around.

\subsubsection{Bias function from abundance matching}
Following the idea of \cite{Sigad2000,Szapudi2004}, a direct way to obtain a mean bias relation is to use the cumulative distribution functions (CDFs), defined as $\mC(\rho)=\int_0^\rho d\rho' \mP(\rho')$, and match their abundances 
\begin{align}
\mC_\rmm(\rho_\rmm)=\mC_\rmt(\rho_\rmt) \,,
\end{align}
such that
\begin{align}
\label{eq:CDFbias}
\rho_\rmm(\rho_\rmt)=\mC_\rmm^{-1}( \mC_\rmt(\rho_\rmt)) \ , \ \rho_\rmt(\rho_\rmm)=\mC_\rmt^{-1}( \mC_\rmm(\rho_\rmm))  \,.
\end{align}
Note that this bias function is built to relate the PDFs of tracer and matter densities by a one-to-one monotonic mapping, which does not assume a local relationship between matter and tracer densities. If there is a large correlation between the matter and its tracer field (for a quantification see the cross-correlation coefficient defined below), this bias function also provides a good fit to a local scatter plot between matter and tracer densities.

\subsubsection{Bias function from conditional mean}
When assuming a local relation between tracer and matter densities, one can also infer mean bias relations from the conditional mean from the scatter plot (SP)
\begin{align}
\label{eq:meanbias}
\rho_\rmm^{\rm SP}(\rho_\rmt):=\langle\rho_\rmm|\rho_\rmt\rangle \ , \ \rho_\rmt^{\rm SP}(\rho_\rmm):=\langle\rho_\rmt|\rho_\rmm\rangle \,.
\end{align}
Note that, in contrast to the bias functions from abundance matching, the composition of the inverse and forward conditional mean bias is not guaranteed to give the identity mapping $\rho_\rmm^{\rm SP}(\rho_\rmt^{\rm SP}(\rho_\rmm))\neq \rho_\rmm$. In particular, this can be a signal for nonlinear bias and a difference in the scatter when fixing matter or tracer density, respectively. We checked that those conditional mean inverse bias functions will be close to the inverse bias function inferred from the CDF method within 3\%. We prefer the CDF method as it is guaranteed to provide a good description for mapping marginal PDFs with conservation of probability.

The (linear) cross-correlation coefficient $r$ between matter and tracer densities in spheres is defined as
\begin{align}
\label{eq:crosscorr}
r = \frac{\langle\rho_\rmm\rho_\rmt\rangle-1}{\sqrt{\langle\rho_\rmm^2\rangle_c\langle\rho_\rmt ^2\rangle_c}} \,.
\end{align}

We show some of the correlation coefficients between matter, halo mass and neutral hydrogen in Table~\ref{tab:corr}. For all redshifts, correlations between matter, neutral hydrogen and mass-weighted halos are all very high and above $95\%$, and neutral hydrogen is almost perfectly correlated with mass-weighted halos for the higher redshifts. Even when comparing matter in real-space to neutral hydrogen in redshift-space, thus absorbing redshift-space distortions in the bias, correlations are still well above $90\%$. This ensures that the bias function from abundance matching, determined purely from the marginal PDFs, will also be a good fit to the scatter plots.

\begin{table}
\centering
\begin{tabular}{l|cccc}
$z$ & $r_{\rmm,\rmHI}$ & $r_{\rmm,\rm{HIz}}$ & $r_{\rmm,\rmhm}$ & $r_{\rmhm,\rmHI}$ \\\hline
1 & 0.969 & 0.918 & 0.982 & 0.945\\
3 & 0.963 & 0.944 & 0.987 & 0.989\\
5 & 0.958 & 0.943 & 0.983 & 0.990\\
\end{tabular}
\caption{Linear cross-correlation coefficients~\eqref{eq:crosscorr} between matter (m), neutral hydrogen in real-space (HI) and redshift-space (HIz) and mass-weighted halos (HM).}
\label{tab:corr}
\end{table}

\subsection{Polynomial bias model in log-densities}
 Following \cite{Jee2012,Uhlemann18bias}, we will use a quadratic model for the (inverse) bias of log-densities $\mu=\log\rho$ in spheres (rather than for the density contrast) which reads
\begin{equation}
\mu_{\rm m} =  \sum_{n=0}^{n_{\rm max}} b_{n}\mu_{\rmt}^{n}\ ,\ n_{\rm max}=2\,.
\label{eq:POLYBIASloginv}
\end{equation}
A heuristic explanation for why a logarithmic transform helps is that it makes the underlying one-point PDFs of matter and biased tracers significantly more Gaussian \citep{Neyrinck09,Carron13}. Hence, it provides a local remapping of nonlinear densities that approximates initial (Lagrangian) densities for which local polynomial bias models are more adequate than for evolved (Eulerian) densities.
As already  emphasized in \cite{Jee2012}, the reason why equation~\eqref{eq:POLYBIASloginv} can be approximated by a linear bias model for the density fluctuations $\delta_{\rmt} = \hat b_1\delta_{\rmm}$ on large scales is that the ranges of log-densities $\mu_\rmt$ and $\mu_\rmm$ become small and not because the bias relation itself becomes linear. This is particularly relevant when focusing  on the tails of the distribution of densities and hence the regime where linear bias is insufficient.
\begin{figure*}
\centering
\includegraphics[width=0.65\columnwidth]{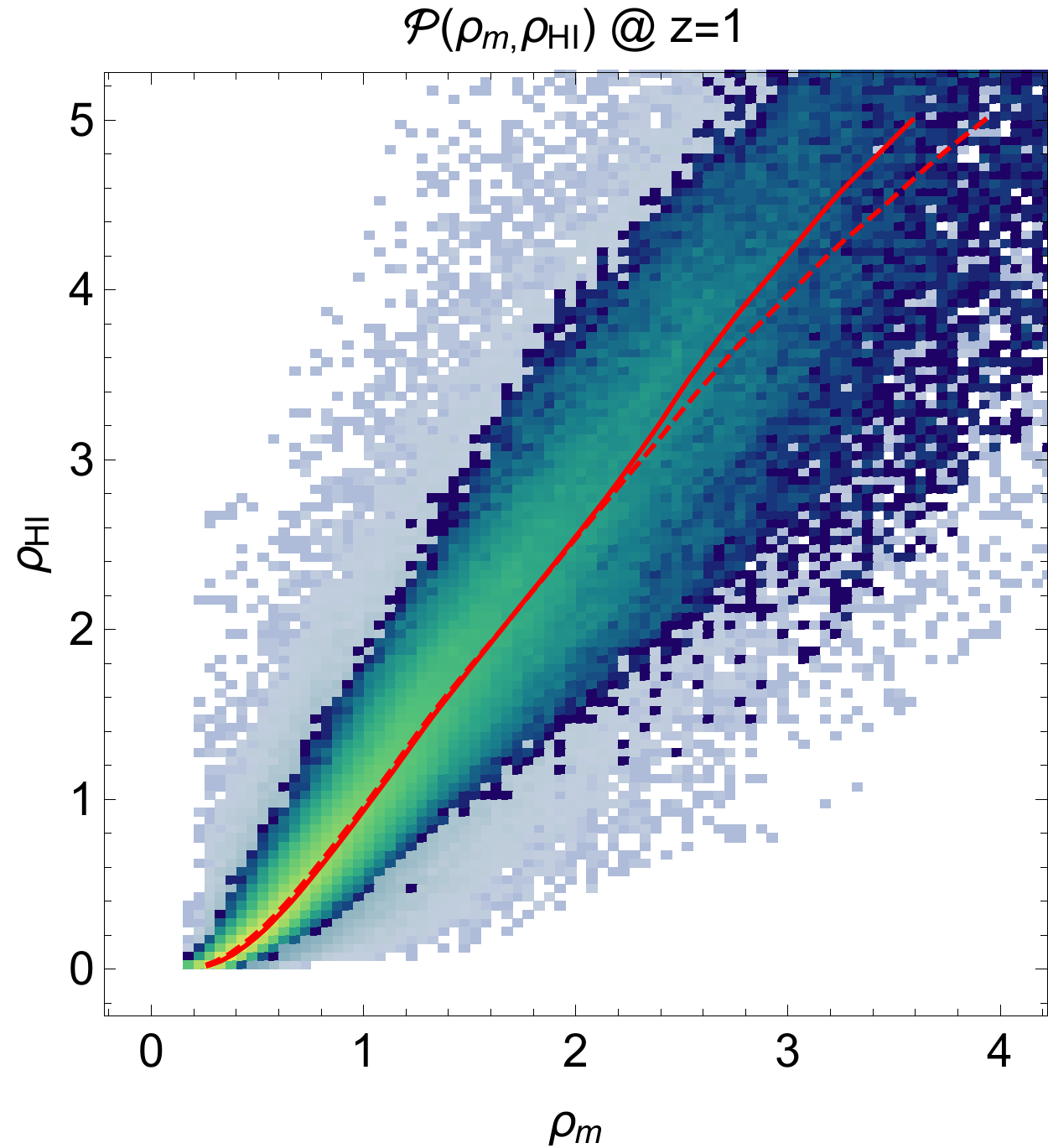}
\includegraphics[width=0.65\columnwidth]{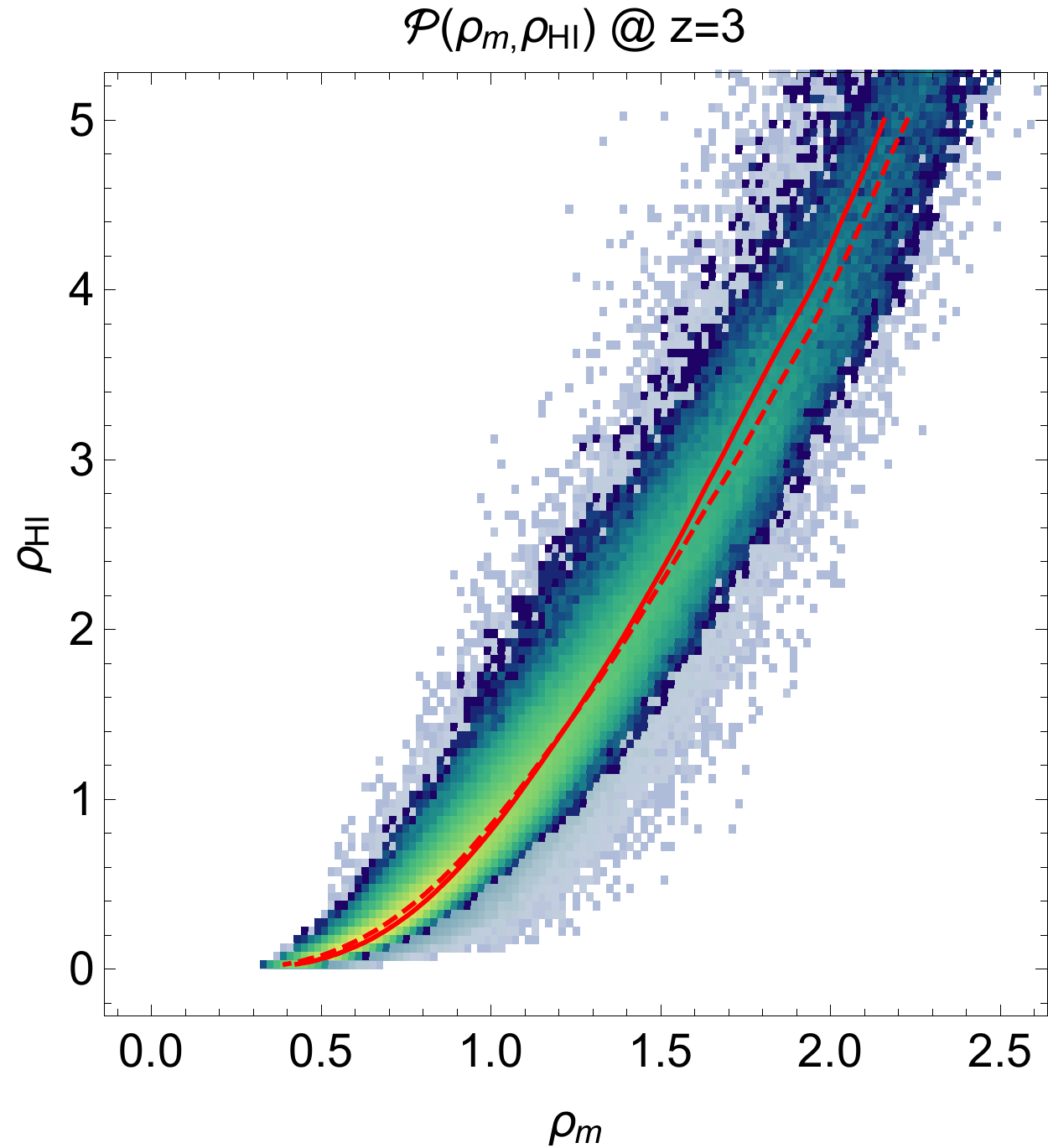}
\includegraphics[width=0.65\columnwidth]{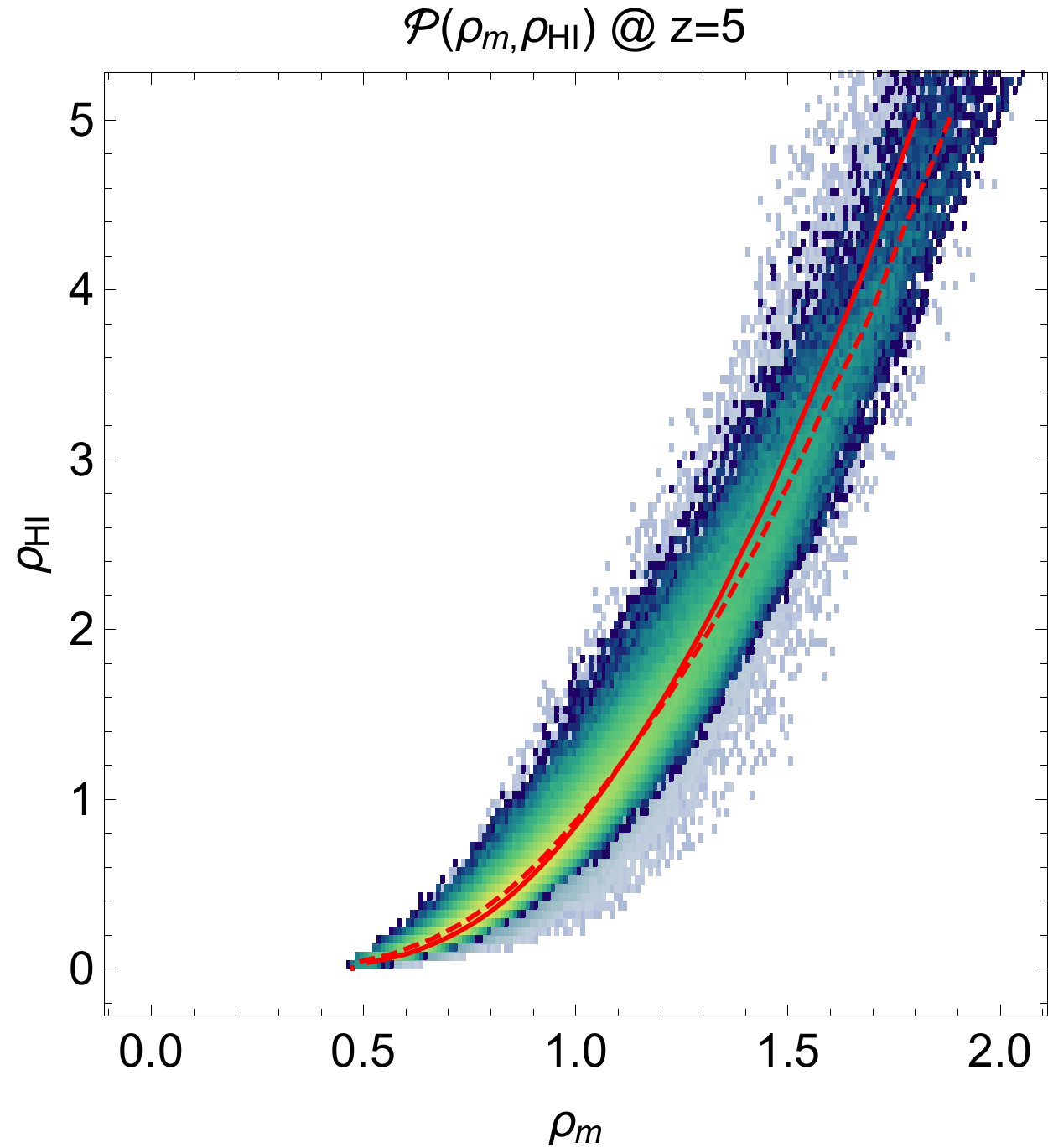}
\caption{Density scatter plots of the neutral hydrogen density $\rho_{\rmHI}$ in real-space (blue-green) and redshift-space (grey) versus the matter density in real-space $\rho_{\rm m}$ for radius $R=5$ Mpc$/h$ at redshifts $z=1,3,5$ (left to right). The figure also shows the parametrisation-independent bias obtained from the CDF in real-space (dashed red line) and redshift-space (solid red line). The main impact of redshift-space distortions is to increase the scatter while the mean bias relation is almost unchanged for average densities and mostly affected in the positive density tails.}
\label{fig:Scatter}
\end{figure*}

\begin{figure*}
\centering
\includegraphics[width=0.65\columnwidth]{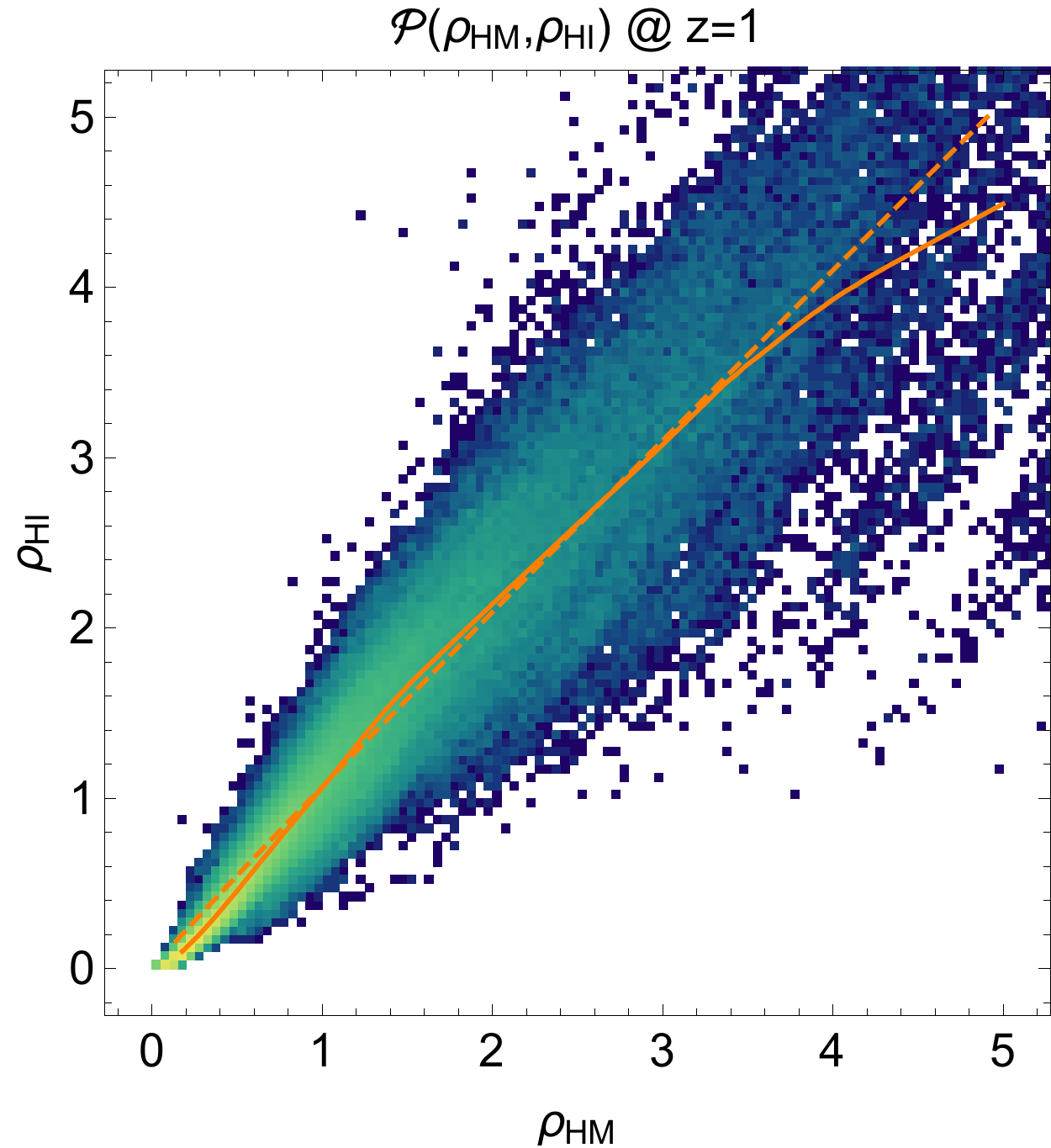}
\includegraphics[width=0.65\columnwidth]{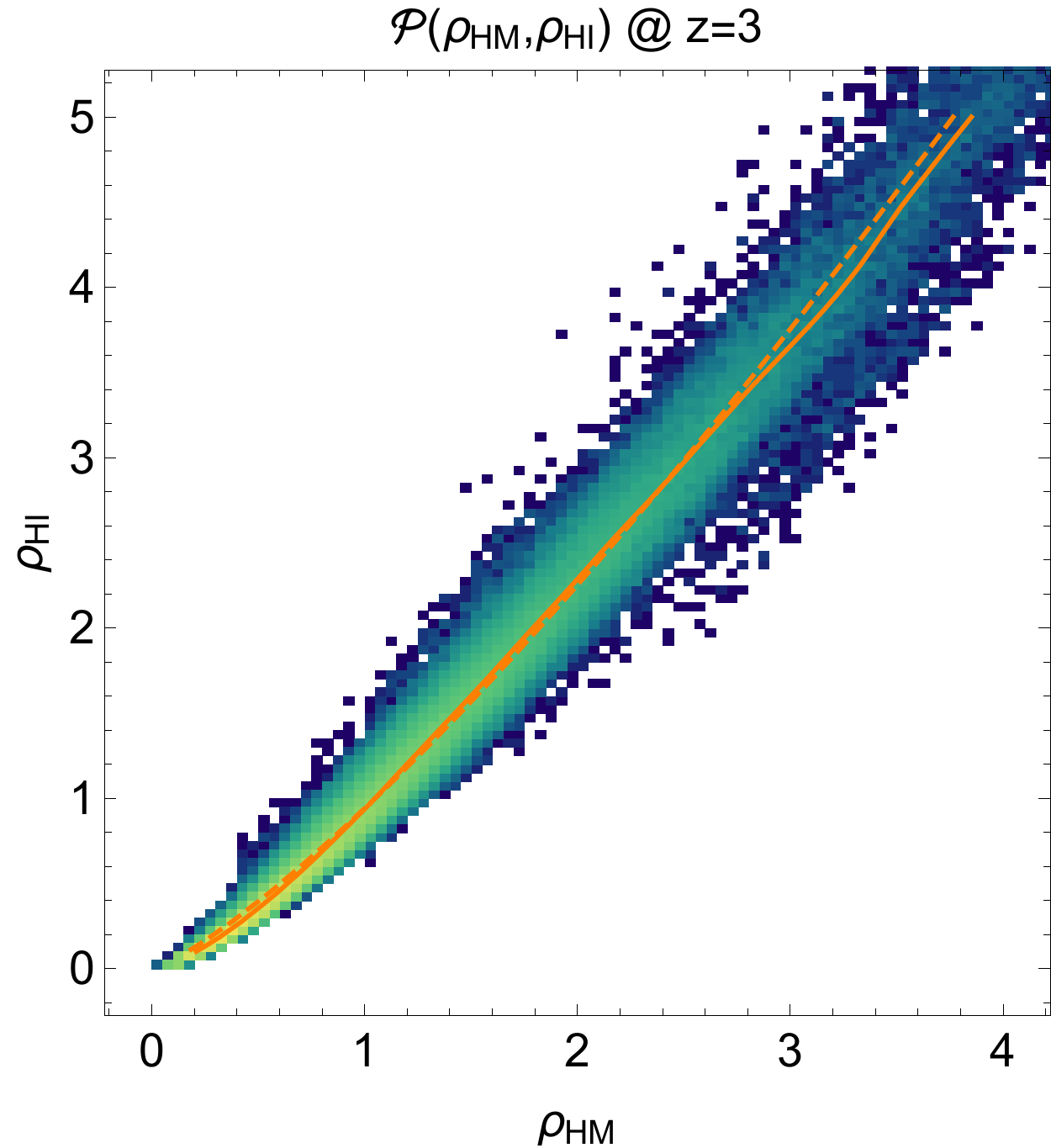}
\includegraphics[width=0.65\columnwidth]{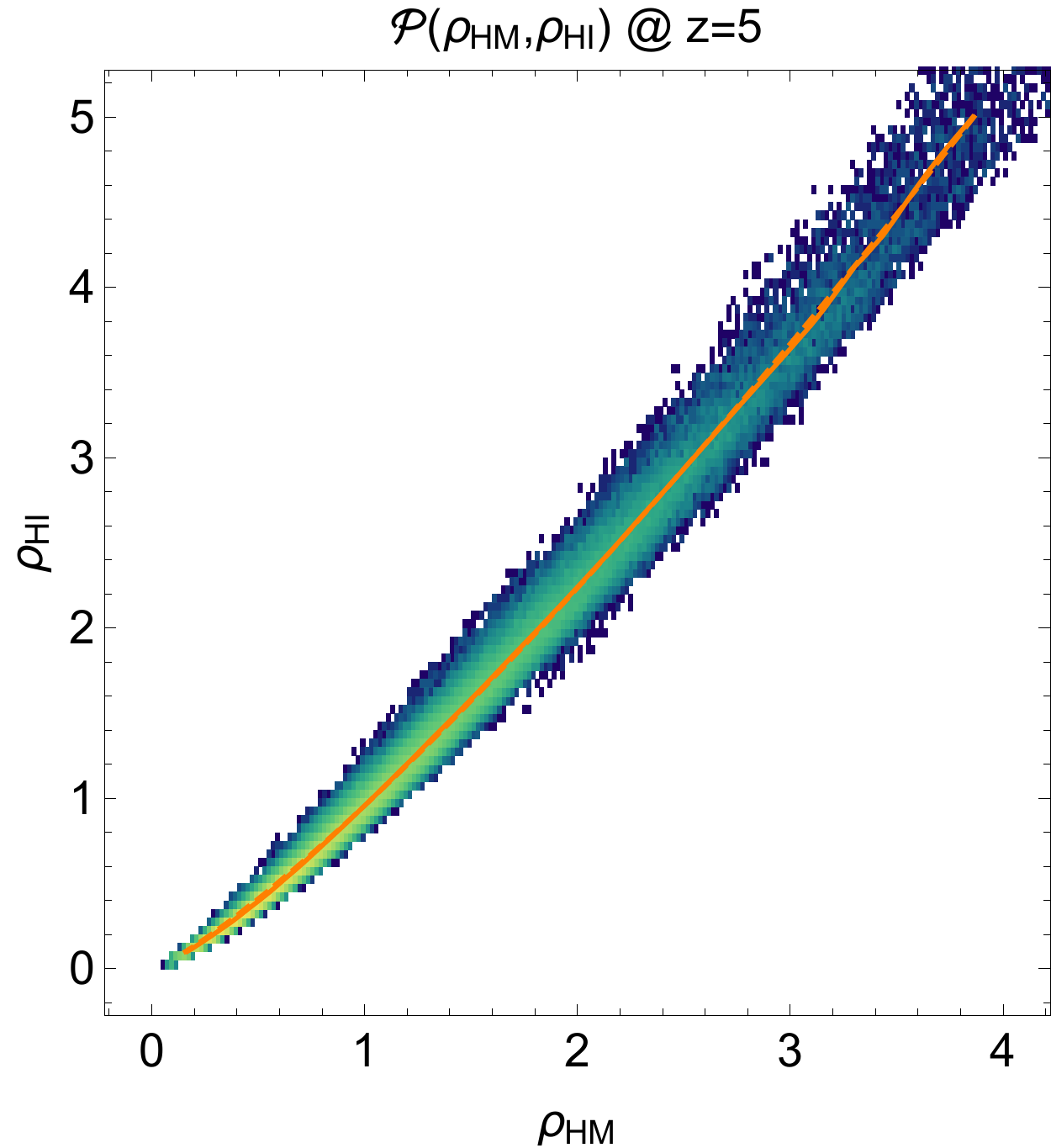}
\caption{Density scatter plots of the real-space neutral hydrogen density $\rho_{\rmHI}$ versus the mass-weighted halo density $\rho_{\rm HM}$ for radius $R=5$ Mpc$/h$ at redshifts $z=1,3,5$ (left to right). The figure also shows the parametrisation-independent bias obtained from the CDF (solid orange line) and a linear fit in log-densities (dashed orange line). This plot shows that the neutral hydrogen closely traces the mass in halos with an almost linear relationship.}
\label{fig:ScatterHIvsHaloMass}
\end{figure*}

\section{Results}
\label{sec:results}
Until recently, the matter-tracer relationship for counts-in-cells has been mainly investigated for dark matter halos and galaxies \citep{Szapudi2004,ManeraGaztanaga11,Desjacques18review,Uhlemann18bias,Friedrich17,Salvador18}. We extend recent results for the relation between neutral hydrogen and matter densities in spheres from \cite{Villaescusa-Navarro18}, finding simple, yet accurate, bias models for their description and use them to predict counts-in-cells statistics for neutral hydrogen.

\subsection{Scatter plots and mean bias models}

\subsubsection{Scatter plots between matter and neutral hydrogen}
Figure~\ref{fig:Scatter}  presents scatter plots comparing densities in spheres for neutral hydrogen $\rho_{\rmHI}$ versus matter densities in real-space $\rho_{\rmm}$ for redshifts $z=1,3,5$ and radius $R=5$ Mpc$/h$. 

We observe that the hydrogen and matter distributions are closely related, as expected from Table~\ref{tab:corr}. Moreover, the scatter around the mean bias (dashed line) increases with cosmic time due to gravitational collapse and astrophysical effects. To make contact with observables, it is required to consider the effect of redshift-space distortions. In Figure~\ref{fig:Scatter} measurements of neutral hydrogen in redshift-space are shown with the shaded points. Redshift-space distortions manifest themselves in an increased variance and are particularly strong at low redshifts where velocity dispersions within halos are typically larger. Moreover, as velocity dispersions become larger in overdense regions, the additional scatter increases with density, and coherent infall into overdensities increases the neutral hydrogen density in redshift-space (solid lines) compared to real-space (dashed lines). While most of the additional scatter when comparing neutral hydrogen densities in redshift space to real-space matter densities comes from the nonlinear mapping from real- to redshift-space, the scatter between redshift-space densities of neutral hydrogen and matter is larger than the scatter between the corresponding real space densities, in particular at higher redshifts.

\subsubsection{Scatter plots between halo mass and neutral hydrogen}
Furthermore, we present a scatter plot between the mass-weighted halo density and neutral hydrogen in Figure~\ref{fig:ScatterHIvsHaloMass}, which shows a much more linear relation (in log-densities) than previously seen for the matter field. This reflects the fact that most of the neutral hydrogen mass is embedded into halos; nearly all mass at $z=0$ and still $90\%$ at $z=5$ \citep{Villaescusa-Navarro18}. Since the amount of neutral hydrogen is sensitive to halo mass, the neutral hydrogen density in spheres is closely related to the mass-weighted halo density in spheres. Our scatter plots of average halo mass and neutral hydrogen in cells complement the halo HI mass functions from Figure~4 of \cite{Villaescusa-Navarro18} showing the HI-halo relation on an object-by-object level. The relation of neutral hydrogen to mass-weighted halo densities is interesting for two reasons. First, mass-weighted halos are in turn closely related to luminosity-weighted galaxies. Hence, joint studies of intensity mapping and galaxy surveys \citep{Bull15,Pourtsidou17} could provide valuable information. In Figure~\ref{fig:halovsHIPDF}, we demonstrate the similarity of mass-weighted halo and neutral hydrogen PDFs by plotting them as a function of the log-density rescaled by the variance. We observe that both neutral hydrogen and halos are nonlinearly biased with respect to the matter, but neutral hydrogen is close to a linearly biased version of halo mass.
Second, the close relation between halo mass and neutral hydrogen density in cells suggests that mocks obtained from populating bound dark matter structures with neutral hydrogen in a halo model approach \citep{Padmanabhan17,Villaescusa-Navarro18,Wolz18} are expected to give accurate results for counts-in-cells on those scales. This could make it possible to study counts-in-cells statistics of neutral hydrogen in larger volumes, taking advantage of cosmological simulations for dark matter.

\subsubsection{Mean bias parametrisations}
\begin{figure}
\includegraphics[width=\columnwidth]{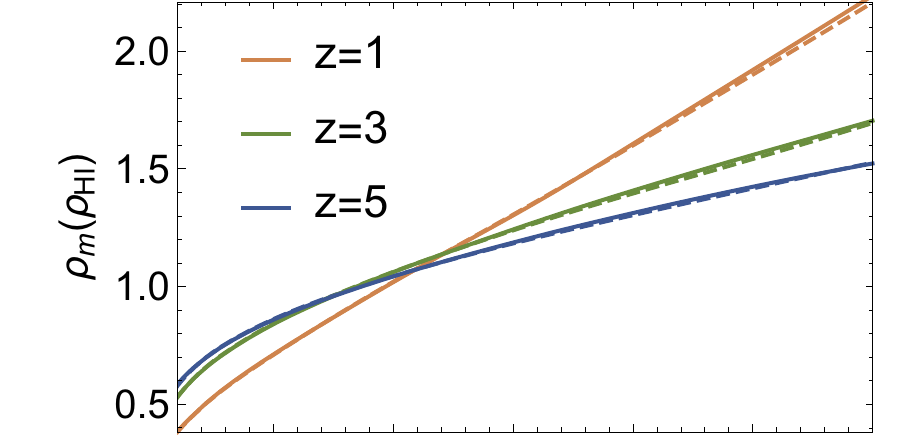}\\
\includegraphics[width=\columnwidth]{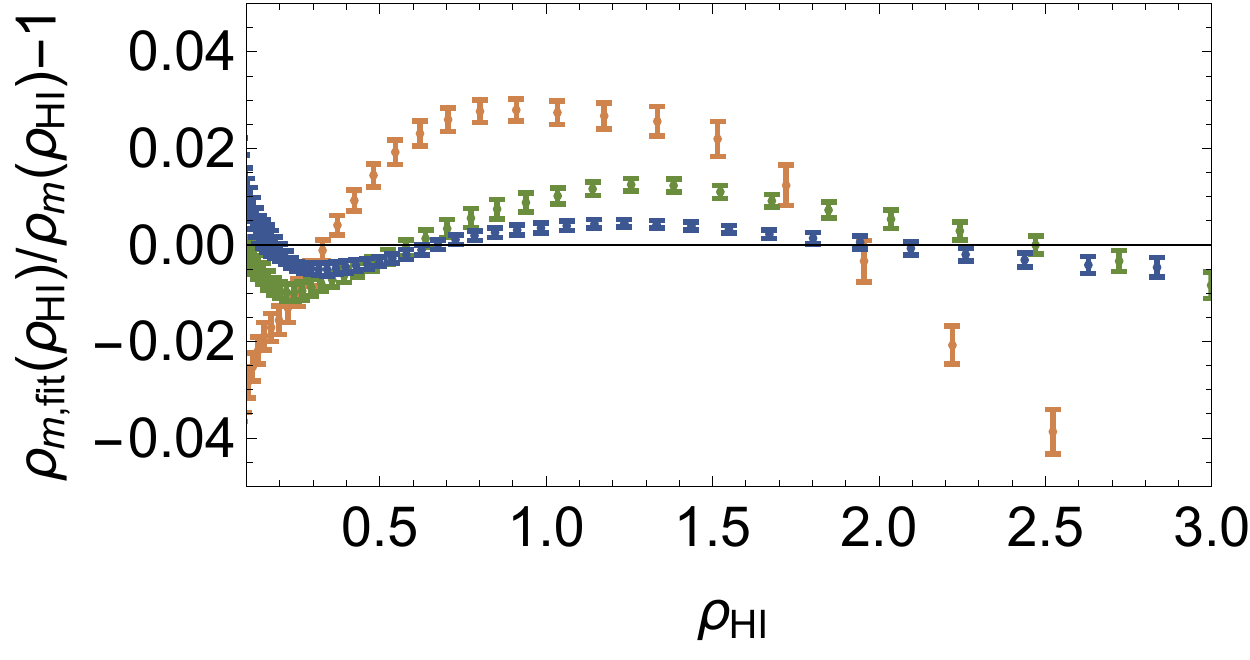}
\caption{{\bf Upper panel:} Parametrisation-independent bias function from the cumulative real space PDFs according to equation~\eqref{eq:CDFbias} using the measured HI PDF with the measured matter PDF (solid lines) or the theoretically predicted matter PDF with the halofit variance (dashed lines). This bias function plots matter densities as function of the neutral hydrogen densities in spheres of radius $R=5$ Mpc$/h$ at redshifts $z=1,3,5$ (orange to blue).
{\bf Lower panel:} Residuals between the parametrisation-independent bias function from the upper panel and a quadratic bias fit in log-densities~\eqref{eq:POLYBIASloginv} in real-space for HI .}
   \label{fig:CDFinvbias}
\end{figure}

Let us now focus on the mean CDF bias functions~\eqref{eq:CDFbias} which are depicted in the scatter plots in Figures~\ref{fig:Scatter}~and~\ref{fig:ScatterHIvsHaloMass}. In Figure~\ref{fig:CDFinvbias}, we show a comparison of the CDF bias obtained from combining the measured neutral hydrogen PDF with either the measured matter PDF, or the fully theoretical PDF model with the halofit variance. We find good agreement at sub-percent level close to the peak and deviations below 5\% for a wide range of densities.
This is encouraging, because it means that the fully theoretical matter PDF, combining \eqref{eq:PDFfromPsi2norm} with the halofit variance, can be used to extract a parametrisation-independent mean bias function from a measured neutral hydrogen PDF. 

Next, we validate our bias parametrisation by comparing the residuals between parametrisation independent CDF-bias~\eqref{eq:CDFbias} and quadratic fits in the log-density according to equation~\eqref{eq:POLYBIASloginv} in Figure~\ref{fig:CDFinvbias}. We fit the parametric bias model \eqref{eq:POLYBIASloginv} to the parameter-independent bias function~\eqref{eq:CDFbias} using the chosen logarithmically spaced sampling points. We estimate the measurement errors at the sampling points using a jackknife estimator (i.e. we derive a CDF bias function for each subset and compute their scatter at all sampling points.). Assuming the measurement errors to be normally distributed and performing an ordinary least squares regression, yields the bias parameters displayed in Table~\ref{tab:biasfit}. As can be seen in Figure~\ref{fig:CDFinvbias}, the quadratic bias model for the logarithmic densities agrees very well with the parametrisation-independent result, both in real-space (solid lines) and redshift-space (dashed lines). We obtain an accuracy  of approximately 4\% at $z=1$ and 1\% at $z=3$ and $5$ over a wide range of densities. When using the fully theoretical matter PDF and determining the bias parameters from those functions), we reproduce the bias parameters from Table~\ref{tab:biasfit} at a few, sub-, 10 percent level for $b_0, b_1, b_2$ respectively. In this case we decided to employ the measurement erros from the former scenario, as due to the high correlation between matter and neutral hydrogen, there are cancellations in the uncertainties which are not captured when using the theoretical matter CDF. When including the next higher order bias parameter $b_3$, one finds that it is typically of order $b_2/10$, but can nevertheless slightly improve the approximation. However, we decided in favour of simplicity and truncated the expansion at second order. 

Note that our bias parameters characterise the inverse relation (matter density as function of tracer density) and in particular our linear (inverse) bias $b_1$ will typically have values around $1/3-1/2$ signalling positive linear forward bias $\tilde b$ around $2-3$ which is in line with previous studies of the bias from the two-point correlation at intermediate redshifts \citep{Villaescusa-Navarro18,Castorina17}.

\begin{table}
\centering
\begin{tabular}{| l | cc | ccc}
 & \multicolumn{2}{c|}{variance} & \multicolumn{3}{c|}{inverse CDF bias}  \\\hline
$z$ &  $\sigma_{\mu,\rmm} $& $\sigma_{\mu,\rmHI} $ & $b_{0}$ & $ b_{1}$ & $ b_{2}$\\\hline
1 & 0.602 & 1.228 & 0.0447 & 0.5716 & 0.0539 \\
2 & 0.438 & 1.121 & 0.0701 & 0.4519 & 0.0432\\
3 & 0.338 & 0.986 & 0.0685 & 0.3791 & 0.0325\\
4 & 0.275 & 0.856 & 0.0524 & 0.3453 &  0.0295\\
5 & 0.231 & 0.798 & 0.0450& 0.3083 & 0.0260\\
\hline
$z$ & $\sigma_{\mu,\rmm}^z $& $\sigma_{\mu,\rmHI}^z $&  $b_{0}^z$ & $ b_{1}^z$ & $ b_{2}^z$\\\hline
1 & 0.705  & 1.311 & 0.0627 & 0.5529 & 0.0534  \\
3 &  0.435 & 1.081 & 0.0740 & 0.3546 & 0.0341 \\
5 &  0.306 & 0.878 & 0.0518 & 0.2829 & 0.0237  \\
\end{tabular}
\caption{Collection of simulation results for $R=5$ Mpc$/h$ at redshifts $z=1$ to $5$. The measured nonlinear variances $\sigma$ of the log-density $\mu=\log\rho$ of both matter ($\rmm$) and neutral hydrogen ($\rmHI$) in real-space (upper part) and redshift-space (lower part) along with the bias parameters obtained from fitting the quadratic model from equation~\eqref{eq:POLYBIASloginv} to the bias function obtained from the CDF according to equation~\eqref{eq:CDFbias}.
}
\label{tab:biasfit}
\end{table} 

\begin{figure}
\includegraphics[width=\columnwidth]{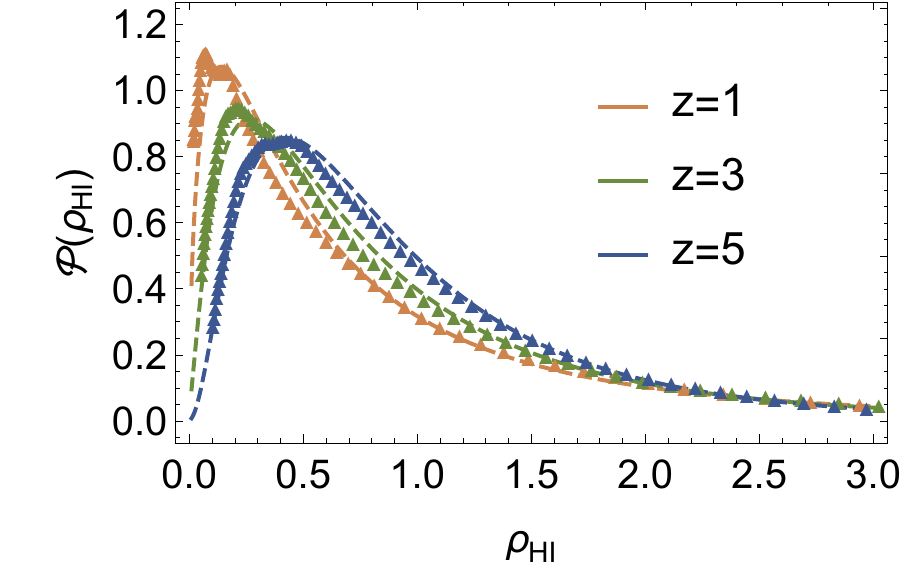}\\
\includegraphics[width=\columnwidth]{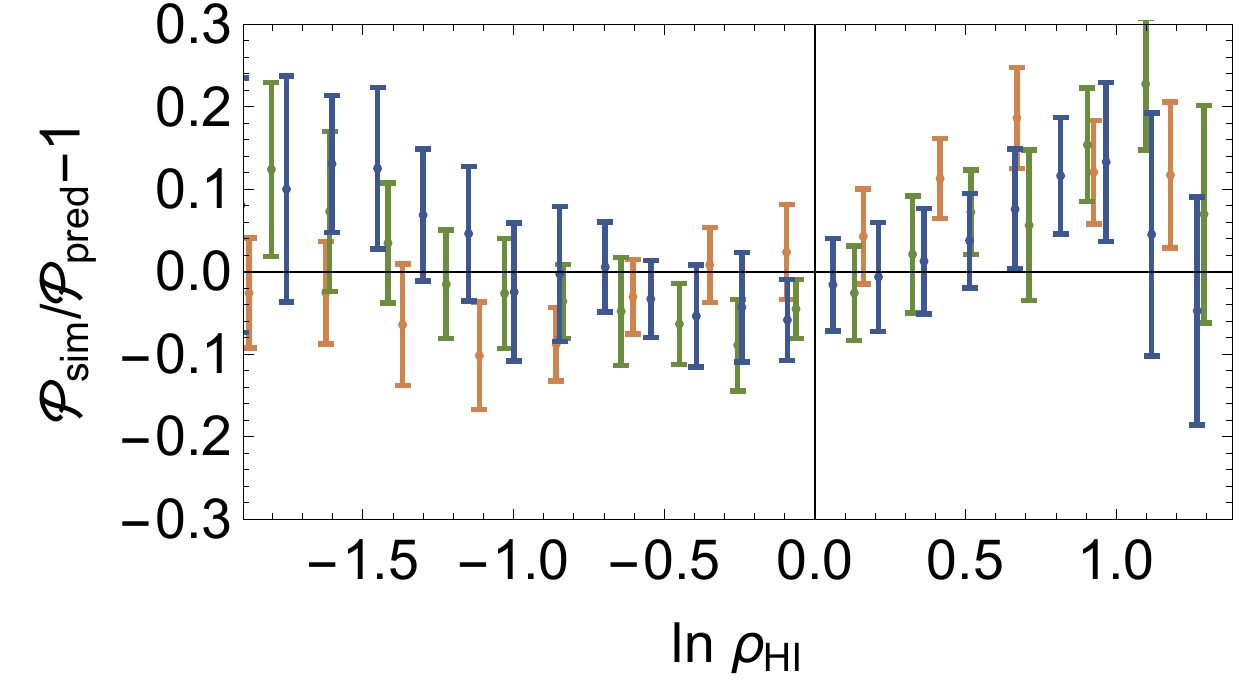}\\
\caption{{\bf Upper panel:} Measured PDF of HI densities in spheres of radius $R=5$ Mpc$/h$ for redshifts $z=1,3,5$ compared to the prediction from large-deviation statistics for matter combined with the quadratic log-bias model (dashed lines) for neutral hydrogen densities in real-space.
{\bf Lower panel:} Residuals between the theoretical predictions and the measured PDFs of HI densities in spheres in real space. }
   \label{fig:HIPDF}
\end{figure}

\begin{figure}
\includegraphics[width=\columnwidth]{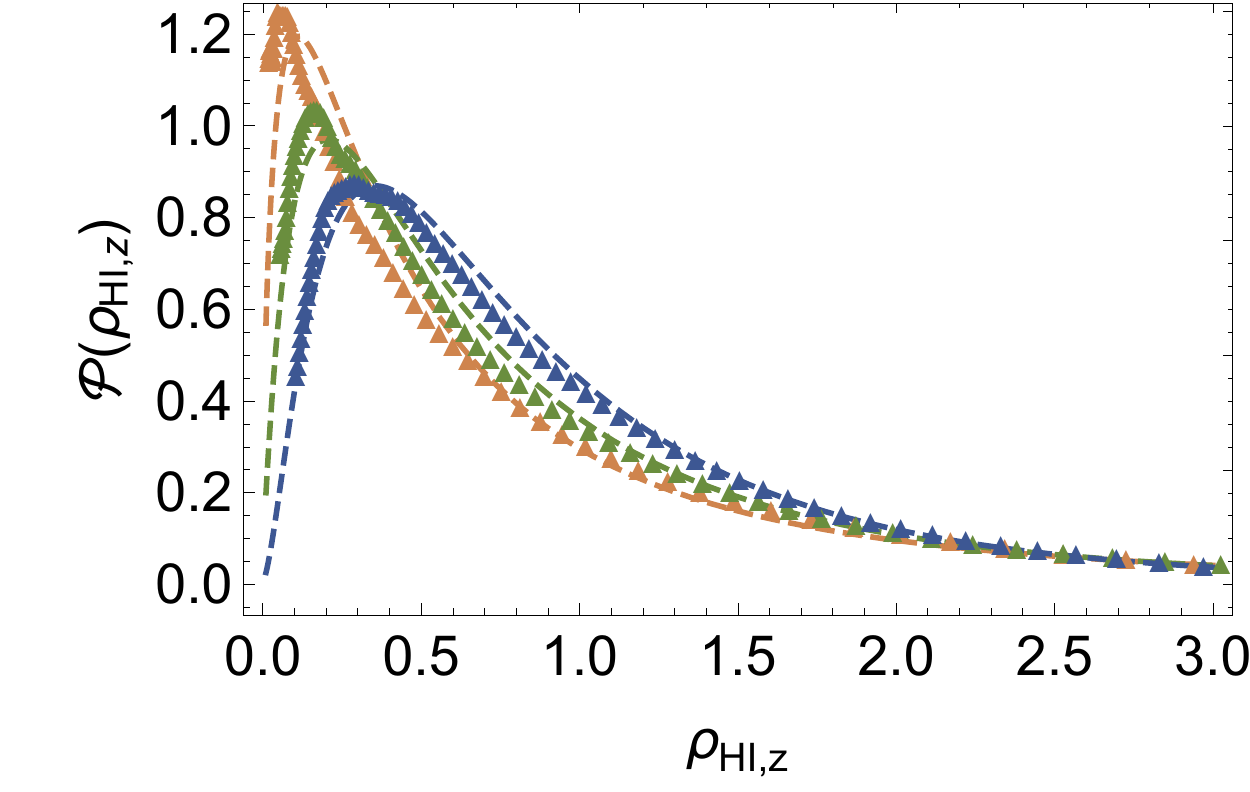}\\
\includegraphics[width=\columnwidth]{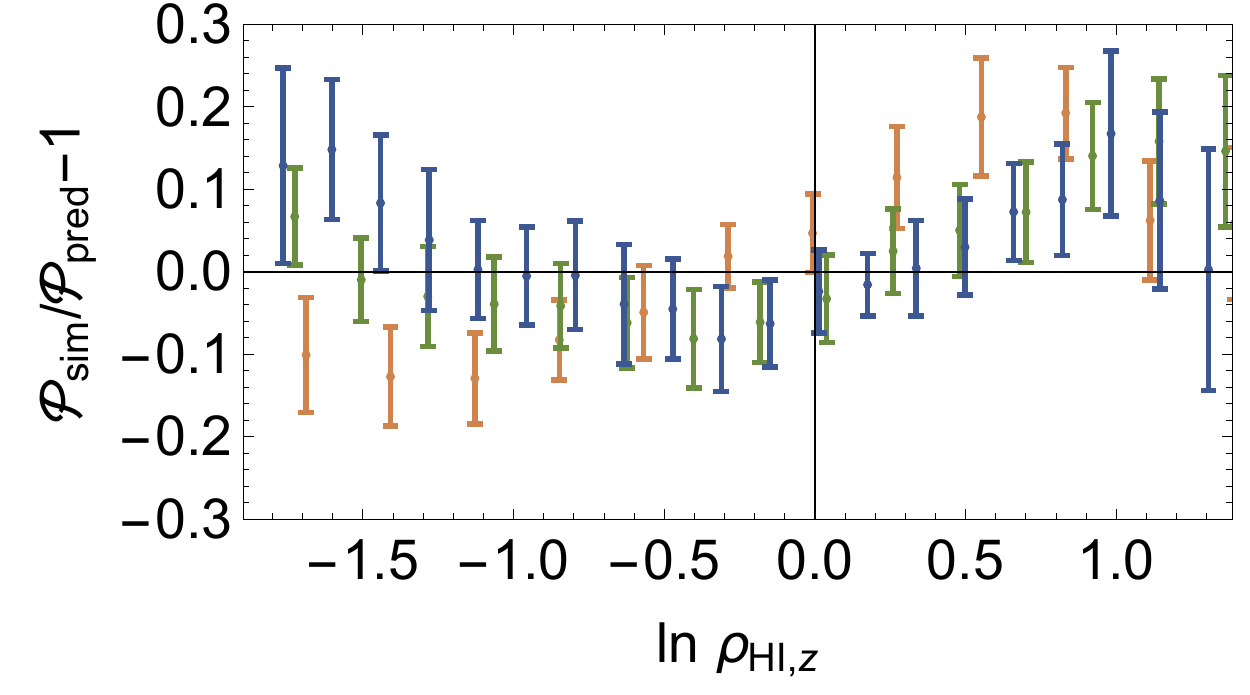}\\
\caption{{\bf Upper panel:} Measured PDF of HI densities in spheres of radius $R=5$ Mpc$/h$ for redshifts $z=1,3,5$ compared to the prediction from large-deviation statistics for matter combined with the quadratic log-bias model (dashed lines) for neutral hydrogen densities in redshift-space.
{\bf Lower panel:} Residuals between the theoretical predictions and the measured PDFs of HI densities in spheres in redshift-space. }
   \label{fig:HIPDFRS}
\end{figure}

\subsection{One-point PDF of neutral hydrogen }
Having established the accuracy of the bias model, let us now combine it with the one-point matter PDF to obtain the one-point neutral hydrogen PDF. The results in real- and redshift-space are shown in comparison to the measurement from the IllustrisTNG simulation in Figures~\ref{fig:HIPDF}~and~\ref{fig:HIPDFRS}, respectively.

In Figure~\ref{fig:HIPDF} we compare our fully predictive theory for matter and the fitted bias model with the measured neutral hydrogen PDF in real space (upper panel). As the variance of the neutral hydrogen density field grows, the amplitude of the PDF tail increases, so the peak of the distribution moves towards lower densities, as they occupy more volume. The residuals (lower panel) show that the theory is able to describe the measurements over a large range in densities at a few percent level. In Figure~\ref{fig:HIPDFRS}, we see that the redshift-space distortions can indeed be incorporated into the bias model, as the measured and predicted PDFs in redshift-space are still in good agreement (upper panel). As expected from the only slightly smaller correlation (cf. Table \ref{tab:corr}), we still see an agreement between theory and the measurement at the few percent level, even though the residuals are slightly larger than for the real-space PDF (lower panel). 

\begin{figure}
\includegraphics[width=\columnwidth]{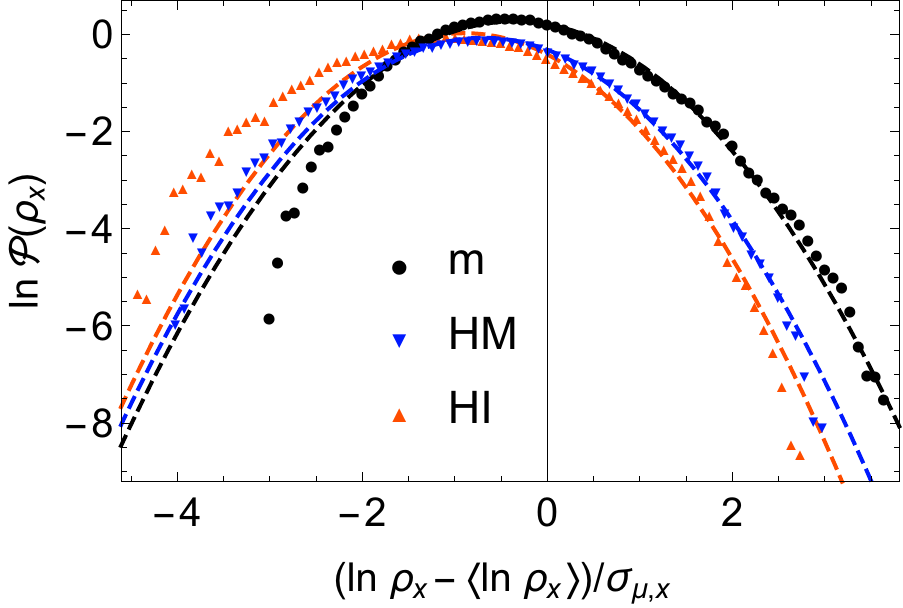}
\caption{Comparison of measured density PDFs of matter (black), mass-weighted halos (blue) and neutral hydrogen (red) in spheres of radius $R=5$ Mpc$/h$ at redshift $z=3$ in real space. The x-axis shows scaled log-densities such that for tracers that are linearly biased in log-densities curves would overlap. Also shown are lognormal fits to the PDFs (dashed lines) which show significant deviations from the measurements for underdensities in matter and neutral hydrogen. }
   \label{fig:halovsHIPDF}
\end{figure}

In Figure~\ref{fig:halovsHIPDF}, we compare the PDFs of neutral hydrogen, mass-weighted halos and matter to the lognormal model \citep{ColesJones91} (dashed lines), which fails in underdense regions and hence renders the large-deviation statistics approach more accurate. We found similar behaviour at redshifts $z=1$ and $z=5$. 

\subsection{Density-dependent clustering of neutral hydrogen}

\begin{figure}
\includegraphics[width=\columnwidth]{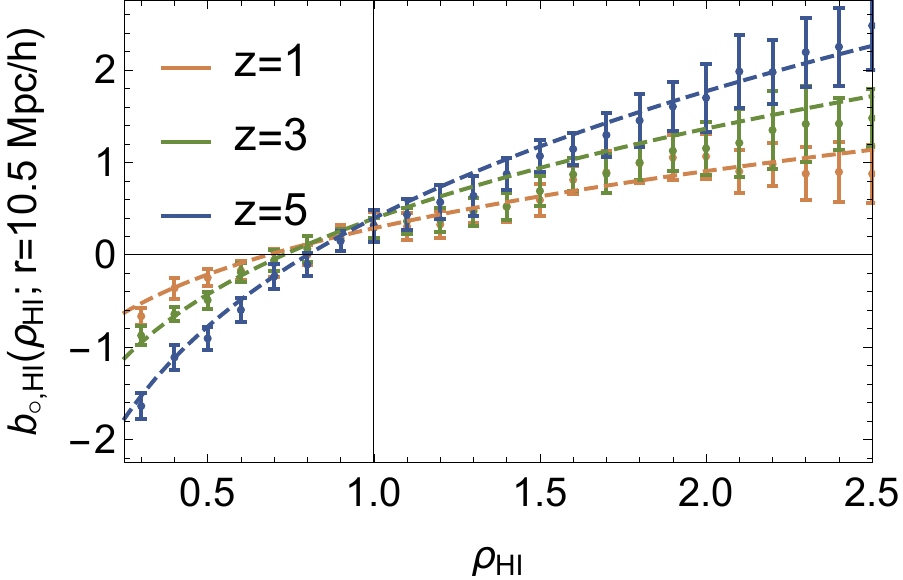}
\caption{Density-dependent clustering encoded in the sphere bias at redshifts  $z=1,3,5$ (orange to blue) for neutral hydrogen in real space as predicted from equation~\eqref{eq:spherebiastracer} (dashed lines) and measured (data points).}
   \label{fig:spherebiasHI}
\end{figure}

The density-dependent clustering signal for biased tracers is interesting, because it offers to break the degeneracy between the nonlinear matter variance and linear bias in the one-point tracer PDF, as demonstrated in \cite{Uhlemann18bias} for the case of halos. Despite the aforementioned limitations of our small-box clustering measurements in IllustrisTNG, the density-dependence of neutral hydrogen clustering in real space displayed in Figure~\ref{fig:spherebiasHI} is clearly a biased version of the density-dependent matter clustering shown in Figure~\ref{fig:spherebiasDM}. The result is in line with the theoretical prediction~\eqref{eq:spherebiastracer} using the bias parameters found for the PDF and an approximation for the ratio of correlation functions $\sqrt{\xi_{\circ,\rmm}/\xi_{\circ,\rmt}}  \simeq \exp(b_0) b_1$ based on a first order expansion of the log-bias model~\eqref{eq:POLYBIASloginv}. The density-dependence of neutral hydrogen clustering is shallower than for dark matter, mainly due to linear bias, but also changes shape due to the nonlinear bias term.

\section{Conclusions and Outlook}
\label{sec:Conclusion}
\textit{Summary.} Building on recent ideas from large-deviation statistics, an accurate theoretical model for counts-in-cells statistics of neutral hydrogen is described. The idea is to rely on analytical predictions for matter and relate them to tracers using a mean bias relation. When combining the analytical results for matter with a nonlinear variance from halofit, one obtains a fully predictive matter PDF that makes it possible to extract a non-parametric bias function from the neutral hydrogen PDF.
Based on measurements in the hydrodynamic simulation IllustrisTNG, we determine the relation between matter and neutral hydrogen densities in spheres of $R=5$ Mpc$/h$ from redshift $z=5$ down to $z=1$. The resulting non-parametric bias relation is well-described by a bias expansion up to second order in log-densities, in line with previous results for halos that host most of the neutral hydrogen. 

The main results for the neutral hydrogen PDF are displayed in Figures~\ref{fig:HIPDF} and~\ref{fig:HIPDFRS} and demonstrate the few percent-level accuracy of the combined analytical model for matter and a mean bias fit both in real and redshift space. In addition, we detect a density-dependent clustering signal for neutral hydrogen (Figure~\ref{fig:spherebiasHI}) that can, in principle, be used to break the degeneracy between the linear tracer bias and the nonlinear variance and jointly constrain $b_1$ and $\sigma_8$. We emphasize that in this paper we are considering single-dish like observations, where neutral hydrogen fluctuations can be directly measured in configuration space. For interferometry observations, the directly observable quantity is the Fourier transform of the intensity flux. Thus, in that case, an approach bearing closer resemblance to observations will be to consider the PDF of mode amplitudes in Fourier-space.

\textit{Fundamental physics.} Future intensity mapping surveys will map gigantic volumes that are ideally suited for counts-in-cells statistics that probe the rare event tails and the growth of structure sensitive to dark energy \citep{Codis2016DE}. The regions of particularly low and high density indeed contain considerable information about fundamental physics such as primordial non-Gaussianity \citep{Uhlemann18pNG} and massive neutrinos \Ref{(Uhlemann \& Villaescusa-Navarro, in preparation)}. 21cm offers a unique technique to observe the 3-dimensional matter density field that allows to go beyond current galaxy surveys, where clustering properties of SDSS galaxy clusters are already used to approach constraints on neutrino mass \citep{Emami17}. We also observed a very close correlation between neutral hydrogen and mass-weighted halo densities, which in turn is expected to translate to luminosity-weighted galaxies. This property could be used for synergies between intensity mapping and redshift galaxy surveys \citep{Bull15,Pourtsidou17}.

\textit{Astrophysics.} Another interesting direction could be to employ the accurate analytical, beyond lognormal model for one-point statistics of dark matter \citep{Uhlemann16log} to probe high-redshift astrophysics. Intensity mapping can be done with lines different from the 21cm spin-flip line of neutral hydrogen, which are sensitive to different astrophysical processes \citep{Suginohara99,Fonseca17} and can probe various environments such as hotter hydrogen gas (Ly$\alpha$), ionised regions (C II) or cool dense molecular gas (CO).
In this context, \cite{Breysse17} introduced the probability distribution of voxel intensities and demonstrated its application to CO emission finding constraints on the luminosity function of the order of 10 percent. In this study, a lognormal matter distribution has been used in combination with a linear relation between halo mass and CO luminosity. The idea is to use the one-point statistics of intensity fluctuations which depend on both the spatial distribution of matter or halos, and also the luminosity function. The luminosity function contains interesting information about the detailed astrophysical conditions within the line emitters, such as star formation rates and metallicities, that can be constrained through the measured PDF of voxel intensity which complements information from the power spectrum \citep{TveitIhle18}.

\textit{Foregrounds.} A serious obstacle for intensity mapping observations is the fact that the amplitude of the galactic and extragalactic foregrounds can be several orders of magnitude higher than the one of the cosmological signal.
Foreground cleaning is thus of pivotal importance and usually takes advantage of the rather smooth frequency spectra of foregrounds that disentangle them from the cosmic signal
which maps the distribution of structures along the line of sight and hence has a significant amount of structure in frequency space. Foregrounds that are constant across the sky are not expected to constitute a serious problem for counts-in-cells statistics, as they just offset the overall mean density. Foregrounds that are spatially varying on scales comparable to the size of the cells would add an extra foreground density fluctuation in every cell, roughly corresponding to extra scatter of the observed intensity around the true density. We have seen before that one-point PDFs are rather robust against scatter such that one has mainly to model the effect of foregrounds on the mean relation. While it is  beyond the scope of the present work, one could quantitatively assess the impact of foregrounds on counts-in-cells statistics through mocks built for 21 cm intensity mapping experiments \citep{Alonso14,Villaescusa-Navarro18}.

\section*{Acknowledgements}
We thank the IllustrisTNG team for providing us with the simulation data for this project. We thank David Alonso for discussions. OL is funded by the Cambridge Trust and an STFC studentship. CU kindly acknowledges funding by the STFC grant RG84196 `Revealing the Structure of the Universe'. The work of FVN and SG is supported by the Simons Foundation through the Flatiron Institute. SC is partially supported by a research grant from Fondation MERAC and by the Programme National Cosmologie et Galaxies (PNCG) of CNRS/INSU with INP and IN2P3, co-funded by CEA and CNES. The IllustrisTNG simulations were run on the HazelHen Cray XC40 super- computer at the High-Performance Computing Center Stuttgart (HLRS) as part of project GCS-ILLU of the Gauss Centre for Supercomputing (GCS). 

\bibliographystyle{mnras}
\bibliography{PDF_HI}



\appendix

\end{document}